\let\final\relax
\let\quantph\relax

\ifx\quantph\undefined
	\documentclass[12pt,a4paper]{scrartcl}
\else
	\documentclass[10pt,a4paper]{scrreprt}
\fi

\newif\iftth

\iftth\else
	\usepackage{scrpage}
	\usepackage{ifthen}
	\usepackage{url}
	\usepackage{amsmath}
	\usepackage[dvips]{graphics,epsfig}
	\usepackage[latin1]{inputenc}
\fi

\def\qpvar#1{\def#1##1{\def#1{##1}}}

\qpvar{\qpno}
\qpvar{\qptitle}
\qpvar{\qpdate}
\qpvar{\qpcontribute}
\qpvar{\qpfirstdate}
\qpvar{\qplastdate}
\qpvar{\qpsolve}

\providecommand{\nothing}{\rule{1em}{0em}--\rule{1em}{0em}}
\iftth
	\newcommand{\bibtitle}[1]{#1}
\else
	\providecommand{\bibtitle}[1]{\emph{#1}}
\fi
\iftth \renewcommand{\emph}[1]{\textbf{#1}} \else \relax \fi
\title{{\normalsize Quantum Information: Problem \qpno}\\ \qptitle}
\author{Institut f\"ur Mathematische Physik\\
	TU Braunschweig, Germany}
\date{Version of \qpdate}
\renewcommand{\maketitle}{

	\pagestyle{empty}
	\thispagestyle{empty}

	\sffamily

	\parbox{\textwidth}{
	\parbox[t]{0.58\textwidth}{\mbox{} \\
	\raggedright\footnotesize
	Open Problems in Quantum Information Theory \\
	Institut f\"ur Mathematische Physik \\
	TU Braunschweig, Germany}
	\hfill
	\parbox[t]{0.4\textwidth}{\mbox{}\\
	\epsfxsize=0.4\textwidth \epsffile{quiprocone_logo.eps}}
	}

	\vfill\vfill

	\begin{center}\begin{minipage}{0.8\textwidth}\begin{flushleft}
	Problem \qpno\\[3ex]
	{\Large\bfseries \qptitle}
	\end{flushleft}

	\vspace{1cm}

	\begin{center}
	\begin{tabular}{|lclc|}
	\hline
	\scriptsize contact: & \qpcontribute & \scriptsize solved by: & \qpsolve \\
	\scriptsize date: & \qpfirstdate & \scriptsize last progress: & \qplastdate \\
	\hline
	\end{tabular} \\[2ex]
	{\scriptsize Version of \qpdate}
	\end{center}
	\end{minipage}\end{center}

	\vfill\vfill\vfill

	\parbox[t]{0.457\textwidth}{\footnotesize For information about the QI open
	problems project at IMaPh refer to the web-pages\\
	\url{http://www.imaph.tu-bs.de/qi/problems/}. Please support us by suggesting
	further interesting problems!}
	\hfill
	\parbox[t]{0.457\textwidth}{\footnotesize For questions, partial results or
	solutions concerning this problem please contact \qpcontribute\ at
	\def\href##1##2{\mbox{\url{##1}}}\qpcontribute.}

	\cleardoublepage
	\pagestyle{myproblems}
	\setcounter{page}{1}
	\normalfont
}
\ifx\quantph\relax
	\let\makemaintitle\maketitle
	\renewcommand{\maketitle}{%
		\chapter{\qptitle}%
 		\begin{center}%
		\sffamily%
 		\begin{tabular}{|lclc|}%
		\hline%
		\rule{0pt}{10.5pt}\scriptsize contact: & \qpcontribute & \scriptsize solved by: & \qpsolve \\%
		\scriptsize date: & \qpfirstdate & \scriptsize last progress: & \qplastdate \\%
		\hline%
 		\end{tabular}%
		\end{center}%
 		\chapterheadendvskip
	}
\fi

\iftth\else
	\ifx\quantph\undefined
		\deftripstyle*{myproblems}{Problem \qpno}{}{\qptitle}{}{\pagemark}{}
		\deftripstyle*{plain}{}{}{}{}{\pagemark}{}
	\else
		\newpagestyle{myproblems}%
			{(0pt,0pt){\headmark}{\headmark}{\headmark}(0pt,0pt)}%
			{(0pt,0pt){\hfill\pagemark\hfill}{\hfill\pagemark\hfill}{\hfill\pagemark\hfill}(0pt,0pt)}%
		\renewpagestyle{plain}%
			{(0pt,0pt){\hfill}{\hfill}{\hfill}(0pt,0pt)}
			{(0pt,0pt){\hfill\pagemark\hfill}{\hfill\pagemark\hfill}{\hfill\pagemark\hfill}(0pt,0pt)}
	\fi
	\global\let\oldsection\section
	\global\def\section#1#{\oldsection*}
\fi

\providecommand{\C}{\mathbf{C}}
\providecommand{\R}{\mathbf{R}}
\providecommand{\imply}{\ensuremath{\Rightarrow}}
\usepackage[dvips]{color}
\usepackage{array}

\qpno{1\,--\,29}
\qptitle{Some Open Problems in Quantum Information Theory}
\qpdate{21 Apr 2005}
\qpcontribute{\href{http://www.imaph.tu-bs.de/qi}{R.\,F. Werner}}
\qpfirstdate{21 Apr 2005}
\qplastdate{\nothing}
\qpsolve{\nothing}

\renewcommand{\nothing}{--}
\definecolor{shadowcolor}{gray}{0.75}
\newcommand{\tocback}[1]{#1}

\setkomafont{pagehead}{\normalfont}
\renewcommand{\bibname}{Literature}

\typearea{default}

\newlength{\cu}
\makeatletter
\def\chapterheadstartvskip{\relax}
\def\chapterheadendvskip{\vspace{\baselineskip}}
\def\@chapter[#1]#2{\ifnum \c@secnumdepth >\m@ne
                        \refstepcounter{chapter}%
                        \typeout{\@chapapp\space\thechapter.}%
						\addtocontents{toc}{\qpno&\protect\multicolumn{4}{l}{\qptitle}%
							&\protect\hyperlink{Outline\thechapter.0}{\thepage}\\%
							&\qpcontribute&\qpfirstdate&\qplastdate&\qpsolve%
							\protect\tabularnewline[2ex]}
						\currentpdfbookmark{\thechapter\ #1}{Outline\thechapter}
                    \else
                      \addcontentsline{toc}{section}{#1}
                    \fi
					\markboth{Problem \qpno\hfill\qptitle}{Problem \qpno\hfill\qptitle}%
					\markright{Problem \qpno\hfill\qptitle}%
                    \addtocontents{lof}{\protect\addvspace{10\p@}}%
                    \addtocontents{lot}{\protect\addvspace{10\p@}}%
                    \if@twocolumn
                      \@topnewpage[\@makechapterhead{#2}]%
                    \else
                      \@makechapterhead{#2}%
                      \@afterheading
                    \fi}
\def\@makechapterhead#1{\chapterheadstartvskip%
  {\size@chapter{\sectfont
    \ifnum \c@secnumdepth >\m@ne {\size@section Problem \thechapter\par\nobreak}\fi%
              {\raggedsection \interlinepenalty \@M #1\par}}
    \nobreak\chapterheadendvskip
  }}
\renewcommand*\bib@heading{%
  \section*{\bibname
  }
  \bibpreamble}
\makeatother

\renewcommand{\makemaintitle}{

	\pagestyle{empty}
	\thispagestyle{empty}

	\sffamily

	\parbox{\textwidth}{
	\parbox[t]{0.58\textwidth}{\mbox{} \\
	\raggedright\footnotesize
	Open Problems in Quantum Information Theory \\
	Institut f\"ur Mathematische Physik \\
	TU Braunschweig, Germany}
	\hfill
	\parbox[t]{0.4\textwidth}{\mbox{}\\
	\epsfxsize=0.2\textwidth
	\hspace*{\fill}
	\epsffile{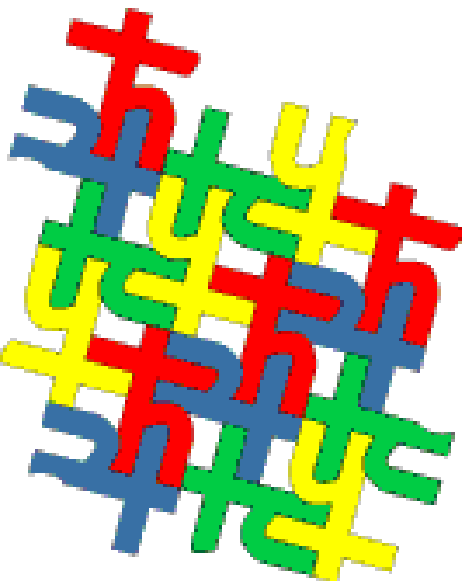}}
	}

	\vfill\vfill

	\begin{center}\begin{minipage}{0.8\textwidth}\begin{flushleft}
	Problems \qpno\\[3ex]
	{\Large\bfseries \qptitle}
	\end{flushleft}

	\vspace{1cm}

	\begin{center}
	\begin{tabular}{|lclc|}
	\hline
	\rule{0pt}{10.5pt}\scriptsize contact: & \qpcontribute & \scriptsize solved by: & \qpsolve \\
	\scriptsize date: & \qpfirstdate & \scriptsize last progress: & \qplastdate \\
	\hline
	\end{tabular} \\[2ex]
	{\scriptsize Version of \qpdate}
	\end{center}
	\end{minipage}\end{center}

	\vfill\vfill\vfill

	\parbox[t]{0.457\textwidth}{\raggedright\footnotesize For information about the QI open
	problems project at IMaPh refer to the web-pages\\
	\url{http://www.imaph.tu-bs.de/qi/problems/}. Please support us by suggesting
	further inter-\\esting problems!}
	\hfill
	\parbox[t]{0.457\textwidth}{\footnotesize For questions,
	additional problems or other con-\\tributions
	please contact \hbox{\qpcontribute} at\\
	\def\href##1##2{\mbox{\url{##1}}}\qpcontribute.}

	\newpage

	\normalfont

	\vspace*{\fill}
	\hspace*{\fill}
	\begin{minipage}{0.8\textwidth}
		\centerline{\textbf{Abstract}}
		This article is a snap-shot of a web site, which has been
		collecting open problems in quantum information for several years,
		and documenting the progress made on these problems. By posting it
		we make the complete collection available in one printout. We also
		hope to draw more attention to this project, inviting every
		researcher in the field to raise to the challenges, but also to
		suggest new problems.

		\vspace{\baselineskip}
		All updates will appear on\\
		\centerline{\url{http://www.imaph.tu-bs.de/qi/problems/}.}
	\end{minipage}
	\hspace*{\fill}
	\vspace*{\fill}
	
	\thispagestyle{empty}
	\pagestyle{myproblems}
}

\usepackage[dvips,a4paper,urlbordercolor={0 0 1},bookmarks=true,pdfpagemode=UseOutlines]{hyperref}
\hypersetup{
	pdftitle={\qptitle},
	pdfauthor={Ole Krüger, Reinhard F. Werner (Editors)},
	pdfkeywords={}
}

\begin{document}

\makemaintitle
\tableofcontents
%\cleardoublepage
\newpage\centerline{}

\addtocontents{toc}{\protect\currentpdfbookmark{Contents}{OutlineTOC}}
\addtocontents{toc}{\protect\begin{tabular}{@{}c@{}c@{}c@{}c@{}>{\protect\centering}p{25\cu}@{}c@{}}}
\addtocontents{toc}{\protect\tocback{No}&\protect\multicolumn{4}{l}{\tocback{Title}}&\protect\tocback{Page}\\%
	&\protect\tocback{Contact}&\protect\tocback{Date}&\protect\tocback{Last progress}&\protect\tocback{Solved by}%
	\protect\tabularnewline[2ex]}
\addtocontents{toc}{\hspace*{5\cu}&\hspace*{27\cu}&\hspace*{18\cu}&\hspace*{18\cu}&\hspace*{27\cu}&\hspace*{5\cu}\\}

\chapter*{Introduction}

\addtocontents{toc}{\protect\currentpdfbookmark{Introduction}{OutlineIntro}}
\addtocontents{toc}{&\protect\multicolumn{4}{l}{Introduction}%
	&\protect\hyperlink{Outline\thechapter.0}{\thepage}%
	\protect\tabularnewline[2ex]}

\markboth{Introduction}{Introduction}%
\markright{Introduction}%

Open problems are among the most important resources of a
researcher. Often enough the key to a scientific discovery is to
ask the right question. But, of course, most of the time things
are not that easy: many problems do resist a serious effort, and
time and again we all come to the point when a problem has
essentially won the fight, and we would be just as happy if
somebody else comes and finally settles it. This is precisely the
sort of problems we want to collect on our problem pages. Of
course, nobody would or should post a problem, for which he or she
has a concrete, promising but untried approach in mind.

The difficulty of the problems in the collection range very
widely, from problems that are, in fact, settled, and turned out
to be easy, to major challenges, which the best people in the
field have struggled with without complete success. A good example
of the first kind is No.~3, which was solved by A. Sudbery,
essentially by pointing out that there is a a Theorem in the
literature which precisely does the job. I count this solution as
a successful example for our page: a call for help that was
answered by someone with a different kind of expertise from the
proposer.

On the other hand, there are the big challenges, like the
additivity problems 7 and 10, shown to be equivalent by Peter
Shor. Probably every quantum information theorist worth his salt
has had a go on that one. Such challenges are landmarks in any
field. If you can make serious progress on one of them, you know
you have really moved.

What makes a list of \emph{open} problems so intriguing is that you
never really know which class any one problem belongs to. Until it
is too late.

Our problem page was created in 1999 and has been growing slowly
over the years. However, it is still not very widely known in the
community. We have therefore obtained the consent of the quant-ph
moderators to post it in its current form. We might post an update
after a couple of years. But you should always look to the site
\url{http://www.imaph.tu-bs.de/qi/problems} for the up-to-date version.

\newpage

\section{Procedures and Policies}

\begin{itemize}
	\item Anybody is invited to contribute problems. They should be
		stated concisely, and in a self-contained way, using only the
		current accepted terminology of the field.
	\item We make an effort to publish problems quickly, but reserve
		the right to reject problems we find less suitable.
	\item We occasionally also post problems that come up in the
		literature and satisfy our criteria.
	\item The best format for submissions is (simple) \LaTeX\ source code, with
		section names taken from a typical published problem (e.\,g. Problem,
		Background, Partial Results, Literature). The source for the whole
		collection is actually maintained in \LaTeX.
	\item Every problem is assigned a \emph{contact person}. This is
		not necessarily the proposer, or the person who formulated the
		problem. However, these colleagues have agreed to keep an eye on
		the problem, are requested to report major partial solutions, and
		will be asked to verify any proposed full solution.
	\item If you want to add a partial solution, or some other relevant remark,
		it is best to send an email both to the contact person and to us. If
		possible, please use \LaTeX\ for this purpose, too.
	\item Full and partial solutions are typically documented via
		citations. If there is no separate paper about the solution, we
		may also post it directly on these pages.
	\item No problem is ever deleted from the list. This is to ensure
		that the entries can be cited in a reliable way. It also helps to
		give due credit to the person who actually solved the problem.
\end{itemize}

%
% Schalter für End-Version: \let\final\relax
% Schalter für Vorschau-Version: \final nicht definiert
%
\ifx\final\relax

%
% Header-Datei laden (voller Pfad, kein Endung ".tex")
%

\ifx\quantph\undefined
\else
% Automatische Definition der Variablen
\qpvar{\qpno}
\qpvar{\qptitle}
\qpvar{\qpdate}
\qpvar{\qpcontribute}
\qpvar{\qpfirstdate}
\qpvar{\qplastdate}
\qpvar{\qpsolve}
\fi

%
% ***************************************************************************
%
% Variablenbelegungen in der LaTeX-Datei,
% einzufuegen durch IMaPh
%
% Diese Variablen sollten NICHT im TEXT-Teil
% definiert oder verwendet werden!
%

% Der folgende Block wird von TTH "verbatim" ausgegeben,
% alle anderen verarbeiten ihn
%%tth: \begin{html}
\qpno{1}
\qptitle{All the Bell Inequalities}
\qpdate{04 Apr 2005}
\qpcontribute{\href{http://www.imaph.tu-bs.de/staff/rfw.html}{R. F. Werner}}
\qpfirstdate{25 Oct 1999}
\qplastdate{22 Oct 2002}
\qpsolve{\nothing}
%%tth: \end{html}

%
% Für Vorschau-Version nur das nötigste einstellen
%
% \nonstopmode kann ggf. auskommentiert werden!
%
\else
\nonstopmode
\documentclass[12pt,a4paper]{article}
\usepackage[latin1]{inputenc}
\markboth{\centerline{\textsl{\bfseries Draft of QI open problem -- content test only!}}}{\centerline{\textsl{\bfseries Draft of QI open problem -- content test only!}}}
\pagestyle{myheadings}
\title{Draft of QI open problem -- content test only!}
\author{No page formatting}
\fi

%
% ***************************************************************************
%
% Praeambel, redigiert durch IMaPh
%

% Keine AMS-Erweiterungen!
% Standardbefehle (\text{}, \R, \C, ...) sind bereits in qiheader.tex
% definiert.

% TTH fuegt Label "BEGINDOCUMENT" ein und ignoriert den Block
%%tth: \iftth BEGINDOCUMENT \begin{document} \else
\ifx\quantph\undefined
	\begin{document}
\fi
\maketitle
% \makeinfo
%%tth: \fi

%
% ***************************************************************************
%
% Text-Teil, redigiert durch IMaPh
%

%
% NICHT VERGESSEN:
%	- ggf. $...$ durch \(...\) ersetzen
%	- Eintraege ins Literaturverzeichnis vollstaendig klammern: \bibitem[ABC]{1:ABC}{A.A.,B.B., ...}
%

\section{Remarks}
The title was taken from a recent exposition by A. Peres \cite{1:Pe}.

\section{Problem}
Find all those linear inequalities
characterizing the existence of joint probability distributions
for all variables in a correlation experiment.

More specifically, suppose that measurements are made on systems,
which are decomposed into \(N\) subsystems. On each of these
subsystems one out of \(M\) observables is measured, producing \(K\)
outcomes each. Thus we consider \(M^N\) different experimental setups,
each of which may lead to \(K^N\) different outcomes, so all in all
\((M K)^N\) probabilities are measured. Classically (in a
``realistic local theory'') these numbers would be
generated by specifying probabilities for each ``classical
configuration'', i.\,e. every assignment of one of the \(K\) values
to each of the \(N M\) observables. Thus the task is to characterize a
convex polyhedron in \((M K)^N\) dimensions (minus a few for
normalization constraints), which is generated by \(K^{(N M)}\)
explicitly known extreme points, in terms of linear inequalities.

For \((N,M,K)=(2,2,2)\) this is solved by the CHSH inequalities. A
general solution for all \(N,M,K\) is highly unlikely to exist.
Therefore we pose the following more managable tasks:

\begin{itemize}

\item Find complete solutions for other small values of \((N,M,K)\).

\item Find efficient ways of generating new inequalities, i.\,e.,
inequalities which cannot be written as convex combinations of
lower order ones.

\item Find infinite families of new inequalities. These could be
complete families of inequalities with certain additional
symmetries.

\item Restrict to ``full correlation functions'', i.\,e., disregard
constraints on marginal distributions. 

\item Do the same for the special case of \emph{correlation
inequalities}. These belong to the case \(K=2\), and are
unchanged, when, for an even number of subsystems, all measurement
outputs are interchanged. Such inequalities are best written in
terms of the expectations of \( A_1 A_2 \ldots A_N\), where each
\(A_i\) takes values \(+1,-1\), resp.
\(-1 \le A_i \le 1\).

\item Decide by what margin these can be violated by quantum states,
or by quantum states with special properties (e.\,g., fixed Hilbert
space dimension, invariance under symmetry transformations or positive partial
transposes). 

\end{itemize}

\section{Background}

This is a special instance of a standard problem in convex
geometry: compute the (maximal) faces of a polyhedron given in
terms of its extreme points. That is: given \(R\) vectors
\(e_k\) in a finite dimensional real vector space,
find the extreme points of the convex set of vectors \(f\) such
that \(f \cdot e_k \le 1\) for all \(k\). By the Bipolar
Theorem \cite{1:Sc} (or ``Farkas' Lemma'', a
special case for polyhedral cones), \(x\) then lies in the
convex hull of the \(e_k\) and the origin, if and
only if \(f \cdot x \le 1\) for all extremal \(f\). It is easy to
decide when such a vector \(f\) is extremal: in that case
\(f\) must be uniquely determined by the equations
\(f \cdot e_k =1\) it satisfies.

To find \emph{some} extreme point is not so difficult: there is a standard
algorithm for maximizing an affine functional on a convex set given in this way
known as the Simplex Algorithm, which runs into an extreme point. It is an entirely
different matter, however, to ask for \emph{all} extreme points. A straightforward
method would be to list all subsets of \(\{1,\ldots,R\}\) with
\((\#\text{elements})=(\#\text{dimensions})\), and to check for each whether the
corresponding set of equations determines an inequality vector \(f\). It is
immediately clear that such a brute force approach to the above problem will end in
an exponential-of-exponential explosion of computing time, and is bound to fail.
There are more intelligent algorithms (e.\,g. the packages available
\href{http://www.netlib.org/voronoi/hull.html}{on netlib},
\href{http://www.ifor.math.ethz.ch/~fukuda/cdd_home/cdd.html}{C++} or
\href{http://www.mathsource.com/Content/Applications/Mathematics/0202-633}{in Mathematica}),
but they, too, all run into serious growth problems for very small
\((N,M,K)\). In fact, there is a theorem by Pitovski to the effect that in a closely
related problem finding the inequalities would also solve some known hard problems
in computational complexity (e.\,g. to the notorious NP = P, resp. NP = coNP questions
\cite{1:Pi}).

So a solution of the problem as posed here necessarily makes
use of the structure of these particular convex sets.

\section{Partial Solutions}
Constraints on the possible range of values of correlations in the
form of inequalities have been investigated for many years (see
the monograph by Frechet \cite{1:Fre}), even before
physicists developed an interest in that subject due to the work
of Bell \cite{1:Be}. The convex geometry aspect of the
above problem was seen clearly by many authors in the last two
decades (e.\,g. \cite{1:Fr}, \cite{1:Ci},
\cite{1:GM}, \cite{1:Pi}, \cite{1:Pe}).
Undoubtedly some
of these have conducted numerical searches for new Bell
inequalities. However, there is only little knowledge about inequalities beyond the
case \((N,M,K)=(2,2,2)\).
Posing this problem is
intended as a focal point for putting together the compilations,
and the existing general observations, so that the state of the
art becomes accessible to a wider community.

\begin{itemize}

\item The first to consider all the possible correlation
functions as a convex set surrounded by the faces of a polyhedron
apparently was M. Froissart \cite{1:Fr}. He identified
these faces with extremal generalizations of Bell's inequalities
and gave some examples up to the case where \((N,M,K)=(2,3,2)\).

\item The case \((2,2,2)\) was analyzed completely by Fine \cite{1:Fi}.
There are only two types of inequalities: one
type just expresses positivity of measured probabilities, the
second is the CHSH-inequality.

\item Tsirelson took up Froissart's idea and concentrated on the
quantum analogue of Bell's inequalities. He pointed out that
quantum theory leads to a convex body wich is in general not a
polytope and thus cannot be described by a finite number of
inequalities. His most complete results were on bipartite
correlation inequalities \((N=K=2)\), where the extremal quantum
correlations are attained by states on Clifford algebras. The
precise structure of the extremal quantum correlations remained
unclear, though. For example, it is not known whether it admits a
description by a finite number of analytic, or even polynomial,
inequalities \cite{1:Ci}.

\item In the work of  work of Garg and Mermin \cite{1:GM}
the case \(K > 2\) was considered, in order to study higher spin
analogues of the standard spin-1/2 situation, and maybe find the
signs of a classical limit. From the point of view of the problem
stated here, the symmetry assumptions of Garg and Mermin  are
rather strong, so that the inequalities obtained describe only a
low dimensional section of the convex body under investigation.

\item Building on \cite{1:GM}, Peres recently claimed
``a graphical method giving a large number of Bell
inequalities of the Clauser-Horne type
\cite{1:Pe}''. Unfortunately, in that paper he merely
applies it to show how to find inequalities for small \((N,M,K)\)
again in larger systems, i.\,e., he does not give any \emph{new}
inequalities in the above technical sense. Peres agrees with
Pitovsky that an algorithm for algebraic construction of these
Farkas vectors runs into serious computational problems unless one
does not use special symmetry properties of these particular
convex sets in order to obtain a more efficient algorithm.

\item Pitowsky and Svozil \cite{1:PS} recently numerically
derived a complete set of inequalities for \((N,M,K)=(3,2,2)\) and \((2,3,2)\)
taking into account constraints on the marginal distributions.
Their results (the coefficients of 53856 inequalities) can be found on their website
(\href{http://tph.tuwien.ac.at/~svozil/publ/ghzbig.html}{\((3,2,2)\)} and
\href{http://tph.tuwien.ac.at/~svozil/publ/3-3.html}{\((2,3,2)\)}).

\item The complete set of correlation inequalities for all \(N\) with
\(M=K=2\) was recently computed by Werner and Wolf \cite{1:WW}. This is somewhat
surprising, since the worst growth of the problem is expected in
the parameter \(N\). There are \(2^{(2^N)}\) inequalities on the
\(2^N\)-dimensional set of correlations corresponding to the maximal faces of a hyper-octahedron,
 which can thus be characterized by a single albeit non-linear inequality. Any of these inequalities
is maximally violated for the generalized GHZ state.
Moreover, one can show that these inequalities are satisfied if
all the partial transposes of the state are positive semi-definite
operators. For the construction and algebraic manipulation of these inequalities a Mathematica 4.0
\href{http://www.imaph.tu-bs.de/ftp/wolf/fourierbell.nb}{notebook} is provided.

\item For \(N=2,M=4\), we get the following
extremal correlation inequalities (\(\mathsf{E}\) stands for
expectation, \(A\) for observables of the first and \(B\) for observables
of the second subsystem):

%-ltoh-	:{}		:\mathsf:<b>:</b>:
%-ltoh-	:comm	:\bigl:::
%-ltoh-	:comm	:\bigr:::

\begin{eqnarray*}
\mathsf{E}\bigl(A_1 (2
B_1+B_2-B_3) + A_4
(B_2+B_3) & + & A_3
(-B_1+B_2-B_3+B_4) \\
& + & A_2
(B_1-B_2+B_3+B_4)\bigr)
\le 6,
\end{eqnarray*}

\begin{eqnarray*}
\mathsf{E}\bigl(A_2 (B_1+ 2 B_2
+B_3-2 B_4) & + & A_4 (2 B_1-2
B_2 +B_3-B_4)\\
& + & A_3 (2
B_1+B_2-2 B_3 +B_4)\\
& + & A_1 (B_1+B_2+2 B_3 +2
B_4)\bigr) \le 10.
\end{eqnarray*}

\item Recently, the relation between the inequalities derived in \cite{1:WW} for
\((N,M,K)=(N,2,2)\)  and distillability has been investigated. It was first
shown by D\"ur \cite{1:Du} that the Mermin-Klyshko inequality can be violated by
multipartite states, which are not \(N\)-partite distillable due to the
positivity of the partial transposes with respect to any \(1|(N-1)\) partition.
For the case of two qubit systems it has then been shown in
\cite{1:Ac,1:ASWa,1:ASWb} that every state violating any \((N,M,K)=(N,2,2)\)
inequality is at least bipartite distillable. It is also proven that there
exists a link between the amount of the Bell inequality violation and the size
of the groups, which have to join in order to be capable of distilling a
multipartite GHZ state. Thus, a strong violation is always sufficient for full
\(N\)-partite distillability.

\item For the case of \((N,M,K)=(2,2,2),(2,3,2)\) the complete set of
correlation inequalities giving the constraint for local hidden variable
models, where one additional bit of classical communication is allowed, has
been constructed in \cite{1:BT}. It is also shown there that quantum theory
satisfies all of these inequalities.

\item Bell inequalities for bipartite systems and more than two outcomes per
observable (and their resistance to noise) have recently been studied in
\cite{1:CGLMP} and \cite{1:MPRG} (see also references therein).

\end{itemize}

Can anyone add to this list?

\ifx\quantph\undefined
	\end{document}
\fi

%
% Schalter für End-Version: \let\final\relax
% Schalter für Vorschau-Version: \final nicht definiert
%
\ifx\final\relax

%
% Header-Datei laden (voller Pfad, kein Endung ".tex")
%

\ifx\quantph\undefined
\else
% Automatische Definition der Variablen
\qpvar{\qpno}
\qpvar{\qptitle}
\qpvar{\qpdate}
\qpvar{\qpcontribute}
\qpvar{\qpfirstdate}
\qpvar{\qplastdate}
\qpvar{\qpsolve}
\fi

%
% ***************************************************************************
%
% Variablenbelegungen in der LaTeX-Datei,
% einzufuegen durch IMaPh
%
% Diese Variablen sollten NICHT im TEXT-Teil
% definiert oder verwendet werden!
%

% Der folgende Block wird von TTH "verbatim" ausgegeben,
% alle anderen verarbeiten ihn
%%tth: \begin{html}
\qpno{2}
\qptitle{Undistillability implies ppt?}
\qpdate{04 Apr 2005}
\qpcontribute{\href{http://www.imaph.tu-bs.de/qi/links.html}{D. Bruß}}
\qpfirstdate{02 Mar 2000}
\qplastdate{25 Oct 2002}
\qpsolve{\nothing}
%%tth: \end{html}

%
% Für Vorschau-Version nur das nötigste einstellen
%
% \nonstopmode kann ggf. auskommentiert werden!
%
\else
\nonstopmode
\documentclass[12pt,a4paper]{article}
\usepackage[latin1]{inputenc}
\markboth{\centerline{\textsl{\bfseries Draft of QI open problem -- content test only!}}}{\centerline{\textsl{\bfseries Draft of QI open problem -- content test only!}}}
\pagestyle{myheadings}
\title{Draft of QI open problem -- content test only!}
\author{No page formatting}
\fi

%
% ***************************************************************************
%
% Praeambel, redigiert durch IMaPh
%

% Keine AMS-Erweiterungen!
% Standardbefehle (\text{}, \R, \C, ...) sind bereits in qiheader.tex
% definiert.

% TTH fuegt Label "BEGINDOCUMENT" ein und ignoriert den Block
%%tth: \iftth BEGINDOCUMENT \begin{document} \else
\ifx\quantph\undefined
	\begin{document}
\fi
\maketitle
% \makeinfo
%%tth: \fi

%
% ***************************************************************************
%
% Text-Teil, redigiert durch IMaPh
%

%
% NICHT VERGESSEN:
%	- ggf. $...$ durch \(...\) ersetzen
%	- Eintraege ins Literaturverzeichnis vollstaendig klammern: \bibitem[ABC]{2:ABC}{A.A.,B.B., ...}
%

\section{Problem}

A state on a bipartite quantum system is called distillable, if from
sufficiently many pairs prepared in that state one can obtain a close
approximation of a maximally entangled singlet state, using only local quantum
operations and classical communication (LOCC). It is well-known that states
with positive partial transpose (PPT) are not distillable. The problem is to
decide the converse.

\section{Background}

This problem has been evident ever since it was shown in \cite{2:HHH1} that
entangled PPT states are undistillable. The two properties, PPT on the one hand
and being undistillable on the other, are mathematically as different as they
can be. Whereas the latter is a variational problem on an unbounded number of
tensor products of density matrices, the first is a simple eigenvalue problem:

\begin{itemize}

\item A bipartite density operator $\rho$ is said to be \emph{PPT} if its
	partial transpose $\rho^{\text{T}_\text{A}}$, defined with respect to some
	product basis via $\langle ij |\rho^{\text{T}_\text{A}}| kl \rangle =
	\langle kj | \rho | il \rangle$, is positive semi-definite, i.\,e., has
	only non-negative eigenvalues.

\item A bipartite state characterized by a density matrix $\rho$ is
	\emph{distillable} if there is a number $n$, such that $\rho^{\otimes n}$
	can locally be projected onto an entangled two qubit state. That is, there
	are two dimensional projectors $Q$ and $P$ acting on the $n$-fold tensor
	product corresponding to Alice respectively Bob, such that

	\[ \left( (P\otimes Q) \rho^{\otimes n} (P\otimes Q)
	\right)^{\text{T}_\text{A}} \]

	has at least one negative eigenvalue. If $n$ copies of $\rho$ have such an
	entangled two qubit subspace, then the state is called
	$n$-\emph{distillable}. There is yet no example of a state, which is
	distillable but not 1-distillable.

\end{itemize}

Using the above criterion of distillability, which was proven by the Horodeckis
in \cite{2:HH}, the problem can be reformulated as \cite{2:DSST}:

\begin{quote}
	Given a completely positive map $S$ such that
	$T S$ is 2-positive (i.\,e. $id_2 \otimes T S$ is
	positive), where $T$ denotes the transpose
	map. Decide whether $T S \otimes T S$ is necessarily
	2-positive.
\end{quote}

\section{Partial Solutions}

\begin{itemize}

\item For special cases like states on Hilbert spaces of dimension $2\times m$
	or Gaussian states it was proven in \cite{2:DCLB}, \cite{2:HHH2} respectively
	\cite{2:GDCZ}, that every such state having a non-positive partial transpose
	(NPPT) is distillable.

\item It was proven in \cite{2:HH}, that every NPPT state can be mapped onto an
	NPPT Werner state by means of LOCC operations. Hence, the matter can be
	decided considering the one-parameter family of Werner states only: if
	there exist any undistillable NPPT states, then there are undistillable
	entangled Werner states.

\item In \cite{2:DCLB}, \cite{2:DSST} numerical evidence has been presented, that
	there may be undistillable NPPT states. Moreover, it was proven
	analytically in \cite{2:DSST}, that for every fixed finite $n$ there is an
	interval of $n$-undistillable entangled Werner states. However, the
	parameter interval for which this statement has been proven, goes to zero
	for $n\rightarrow\infty$.

\item It was proven in \cite{2:EVWW} that if one enlarges the class of allowed
	operations from LOCC to PPT preserving maps, then every NPPT state becomes
	1-distillable. For a proof using entanglemet witnesses and the discussion
	of the tripartite case see \cite{2:KLC}. Note that every PPT-preserving map
	can stochastically be implemented as LOCC operation with an additional PPT
	entangled state as a resource \cite{2:CDKL}.

\item If an additional entangled PPT state $\sigma$ makes an NPPT state $\rho$,
	which is not 1-distillable itself, become 1-distillable, then we say that
	$\sigma$ \emph{activates} the distillability of $\rho$. It has been proven
	in \cite{2:VW} that there are PPT states $\sigma$, which are capable of
	activating every NPPT state. Moreover, the required amount of entanglement
	(measured in terms of any entanglement measure, which is continuous at the
	separable boundary) has been shown to be infinitesimally small \cite{2:VW}.
	In \cite{2:KLC} a formalism was introduced that connects entanglement
	witnesses and the activation properties of a state. Here it was shown that
	there exist three--partite NPPT states with the property that two copies
	can neither be distilled, nor activated.

\end{itemize}

\ifx\quantph\undefined
	\end{document}
\fi

%\newpage\centerline{}

%
% Schalter für End-Version: \let\final\relax
% Schalter für Vorschau-Version: \final nicht definiert
%
\ifx\final\relax

%
% Header-Datei laden (voller Pfad, kein Endung ".tex")
%

\ifx\quantph\undefined
\else
% Automatische Definition der Variablen
\qpvar{\qpno}
\qpvar{\qptitle}
\qpvar{\qpdate}
\qpvar{\qpcontribute}
\qpvar{\qpfirstdate}
\qpvar{\qplastdate}
\qpvar{\qpsolve}
\fi

%
% ***************************************************************************
%
% Variablenbelegungen in der LaTeX-Datei,
% einzufuegen durch IMaPh
%
% Diese Variablen sollten NICHT im TEXT-Teil
% definiert oder verwendet werden!
%

% Der folgende Block wird von TTH "verbatim" ausgegeben,
% alle anderen verarbeiten ihn
%%tth: \begin{html}
\qpno{3}
\qptitle{Polynomial entanglement invariants}
\qpdate{04 Apr 2005}
\qpcontribute{\href{http://www.imaph.tu-bs.de/staff/rfw.html}{R. F. Werner}}
\qpfirstdate{13 Oct 2000}
\qplastdate{18 Dec 2001}
\qpsolve{A. Sudbery}
%%tth: \end{html}

%
% Für Vorschau-Version nur das nötigste einstellen
%
% \nonstopmode kann ggf. auskommentiert werden!
%
\else
\nonstopmode
\documentclass[12pt,a4paper]{article}
\usepackage[latin1]{inputenc}
\markboth{\centerline{\textsl{\bfseries Draft of QI open problem -- content test only!}}}{\centerline{\textsl{\bfseries Draft of QI open problem -- content test only!}}}
\pagestyle{myheadings}
\title{Draft of QI open problem -- content test only!}
\author{No page formatting}
\fi

%
% ***************************************************************************
%
% Praeambel, redigiert durch IMaPh
%

% Keine AMS-Erweiterungen!
% Standardbefehle (\text{}, \R, \C) sind bereits in qiheader.tex
% definiert.
\providecommand{\GL}{\mathrm{GL}}

% TTH fuegt Label "BEGINDOCUMENT" ein und ignoriert den Block
%%tth: \iftth BEGINDOCUMENT \begin{document} \else
\ifx\quantph\undefined
	\begin{document}
\fi
\maketitle
% \makeinfo
%%tth: \fi

%
% ***************************************************************************
%
% Text-Teil, redigiert durch IMaPh
%

%
% NICHT VERGESSEN:
%	- ggf. $...$ durch \(...\) ersetzen
%	- Eintraege ins Literaturverzeichnis vollstaendig klammern: \bibitem[ABC]{3:ABC}{A.A.,B.B., ...}
%

\section{Problem}

We say that two bipartite quantum states $\rho$ and $\sigma$ are
``equally entangled'' if they differ only by a choice of
bases in Alice's and Bob's subspaces, i.\,e., if we can find
unitaries $U_A$, $U_B$, such that

\[ \rho=(U_A\otimes U_B) \sigma (U_A\otimes U_B)^*. \]

An \emph{entanglement invariant} is by definition any real valued
function on the space of bipartite density operators, which
assigns the same value to equally entangled density operators. A
\emph{polynomial invariant} is an entanglement invariant, which
can be computed as a polynomial in the matrix elements of $\rho$.
Note that because we only consider hermitian operators, allowing
polynomials in the matrix elements and their complex conjugates
does not enlarge this class.

The basic problem is to decide the following question:
\begin{itemize}
\item Are the polynomial entanglement invariants \emph{complete}?,\\
	i.\,e., if all polynomial invariants of $\rho$ and $\sigma$ agree, can
	we infer the existence of unitaries $U_A$, $U_B$
	satisfying the above equation?
\end{itemize}

But we may add some further, closely related problems:
\begin{itemize}
\item Given the dimensions of Alices's and Bob's Hilbert spaces,
	name a finite set of invariants which is already complete.
\item Do all this for multi-partite states. In this case even the
	case of pure states is not obvious.
\item Decide whether the set of separable states can be described
	in terms of a polynomial invariant $f$, such that
	$f(\rho)\ge 0$ is equivalent to separability. There are
	many weaker versions of this statement, which may be of interest.
	For example, we might merely ask for a sufficient or a necessary
	separability criterion, and we might allow $f$ to depend on
	the dimensions.
\end{itemize}

\section{Background}

\begin{enumerate}
\item When $d_1$ and $d_2$ are the dimensions of Alice's and Bob's Hilbert
space, respectively, the state space is $(d_1\, d_2)^2-1$ dimensional. Since
phases for $U_A$, $U_B$ drop out of the transformation equation, we may fix
their determinant to be 1, and hence get $(d_1^2-1)+(d_1^2-1)$ for the
dimension of the symmetry group. Subtracting we get an expected manifold
dimension of $(d_1^2-1)(d_1^2-1)$ for the quotient manifold, i.\,e., the manifold
of all invariants. Of course, it may happen that no set of this many
differentiable invariants is sufficient to pin down each equivalence class
uniquely, and more invariants are needed to rule out some discrete choices.
\item It is perhaps useful to recall the ``unipartite'' version of this problem,
i.\,e., the characterization of density operators up to unitary equivalence. The
well-known complete set of invariants in that case is the spectrum of the
density operator (including multiplicities). The eigenvalues are not
polynomial, but the coefficients of the characteristic polynomial (i.\,e., the
elementary symmetric functions of the eigenvalues) or, equivalently the numbers
$a_n=tr(\rho^n)$ are polynomial, and from these the eigenvalues can be
determined. Hence a complete set of invariants are the $a_n$ for
$n=1,...,\text{dimension}$.
\item A basis for the ring of of invariant polynomials (even in the
multi-partite case, and for arbitrary Hilbert space dimensions) was given in
\cite{3:GRB} and \cite{3:R}. Note that any homogeneous polynomial of degree $k$ in
$\rho$ can be written as an expectation value of the $k^\text{th}$ tensor power
of $\rho$, i.\,e., as $tr(\rho^{\otimes k} X)$, with a uniquely determined $X$.
For an $n$-partite system this is an operator on a tensor product of $n\, k$
Hilbert spaces. Invariance requires that it commutes with all unitaries of the
form $U_1^{\otimes k}\otimes ...\otimes U_n^{\otimes k}$, where  $U_m$ is an
operator on the Hilbert space of the $m^\text{th}$ type of systems
($m=1,...,n$). Then the commutation theorem of von Neumann algebras, and the
corresponding result for $n=1$, imply that $X$ must be a tensor product of $n$
permutation operators, each one permuting the $k$ tensor factors belong to one
of the $n$ system types.
\end{enumerate}

\section{Partial Solutions}

\begin{itemize}
\item Y. Makhlin \cite{3:M} has shown completeness in the bipartite
	qubit case. Moreover, he has identified a set of 18 invariants, which is
	sufficient in that case, and has shown that none of these may be omitted
	without destroying completeness.
\item A. Sudbery \cite{3:S} has solved the case of pure three qubit
	states, finding 8 polynomial invariants (6 being the dimension of
	the manifold of all invariants).
\end{itemize}

\section{Solution}

The basic question of principle (are the polynomial entanglement
invariants complete?) is answered in Onishchik and Vinberg's book
\bibtitle{Lie Groups and Algebraic Groups}, which contains the theorem
\cite{3:OV}

\begin{quote}
The orbits of a compact linear group acting in a real vector space 
are separated by the polynomial invariants.
\end{quote}

In other words (those of quantum information theory), if two
states of a multipartite system
are not related by local unitary transformations, then they have
different values for some polynomial entanglement invariant.

It follows that the space of entanglement types of states, i.\,e.\ the
space of orbits factored by normalisation, can be identified with the space
of polynomial
invariants (more precisely, the ring of polynomial functions on this space is 
isomorphic to the ring of polynomial invariants). The dimension of this
space is known in full generality for pure states \cite{3:CHS}. For
two parties it is one less than the dimension of the smaller state
space (a complete set of invariants is the set of Schmidt coefficients,
which sum to 1 by normalisation). For $n>2$, 
if the parties have state spaces with dimensions
$d_1,\ldots,d_n$ in increasing order, then the space of orbits of
normalised states has
dimension 
\[ D_{\text{pure}} = 2\prod_{r=1}^n d_r - \sum_{r=1}^n d_r^2 + n - 2 +
\Delta^2 \]
where $\Delta = d_n - d_1 \ldots d_{n-1}$ if this is positive, otherwise
$\Delta = 0$. If all the parties are qudits ($d_1 = \cdots = d_n = d$) this
becomes 
\[ D_{\text{pure}} = 2d^n - nd^2 + n - 2 .\]
The corresponding dimension for mixed states is
\[ D_{\text{mixed}} = d^{2n} - nd^2 + n - 1 \]
which is probably correct, though a careful treatment has never appeared
in the literature. The general case for mixed states has not been
discussed.

The number of invariants needed to uniquely specify a state up to local
unitary transformations is not the same as the dimension $D$ of the
space of entanglement types; this is in general a curved space, with
complicated geometry. Makhlin's work \cite{3:M} shows that the space
of entanglement types of mixed states of two qubits is a
nine-dimensional manifold in $\R^{18}$ (the ring of polynomial
invariants has 18 generators subject to 9 relations). For pure states of
three qubits, which have $D=6$ (including the norm), a complete set of 
invariants \cite{3:AAJT} consists of
the six independent invariants given in \cite{3:S} together with
one more found by Grassl. Thus the space of orbits of non-normalised
state vectors is a hypersurface in $\R^7$; normalising, the space of
entanglement types of pure
states of three qubits is a hypersurface in real projective 6-space.

The above theorem was used by Hilary Carteret and myself in our proof
\cite{3:CS} that on an 
orbit whose dimension is exceptionally low,
some entanglement invariant has an extreme value. We classified these
exceptional
orbits for pure states of three qubits. 

The condition that the group should be compact is essential, as is shown
by the example of the general linear group GL$(n,\C)$ acting on
$n\times n$ complex matrices by the similarity transformation $X\mapsto
GXG^{-1}$ where
$G\in$ GL$(n,\C)$. The polynomial invariants here are the coefficients in the
characteristic equation of $X$, so two matrices have the same values of
the invariants if and only if they have the same eigenvalues. But having
the same eigenvalues is not sufficient for two matrices to be similar;
if some of the eigenvalues are repeated, there are different possible
Jordan normal forms which are not related by similarity. 

An even simpler example, and one which is relevant to quantum
information theory, is the action of $\GL (m,\C)\times\GL (n,\C)$ on
$m\times n$ matrices by $X\mapsto PXQ^T$ where $P\in \GL (m,\C)$ and
$Q\in \GL(n,\C)$. In
this case there are no polynomial invariants, but matrices can
only be transformed into each other by such a transformation if they
have the same rank. (The rank is a non-polynomial invariant.) If we take
$X$ to be an element of $\C^m\otimes\C^n$ representing a pure state of a
bipartite system, two states are related by this action if there are local
operations which will convert them into each other with non-zero probability.
This generalises the deterministic (unitary) local operations which
define equally entangled states in the statement of the problem. The
corresponding orbits for three qubits have been determined
by D\"ur, Vidal and Cirac \cite{3:DVC}, and for 
four qubits by Verstraete, Dehaene, De Moor and Verschelde
\cite{3:MVDV}.

\ifx\quantph\undefined
	\end{document}
\fi

%\newpage\centerline{}

%
% Schalter für End-Version: \let\final\relax
% Schalter für Vorschau-Version: \final nicht definiert
%
\ifx\final\relax

%
% Header-Datei laden (voller Pfad, kein Endung ".tex")
%

\ifx\quantph\undefined
\else
% Automatische Definition der Variablen
\qpvar{\qpno}
\qpvar{\qptitle}
\qpvar{\qpdate}
\qpvar{\qpcontribute}
\qpvar{\qpfirstdate}
\qpvar{\qplastdate}
\qpvar{\qpsolve}
\fi

%
% ***************************************************************************
%
% Variablenbelegungen in der LaTeX-Datei,
% einzufuegen durch IMaPh
%
% Diese Variablen sollten NICHT im TEXT-Teil
% definiert oder verwendet werden!
%

% Der folgende Block wird von TTH "verbatim" ausgegeben,
% alle anderen verarbeiten ihn
%%tth: \begin{html}
\qpno{4}
\qptitle{Catalytic majorization}
\qpdate{04 Apr 2005}
\qpcontribute{\href{http://www.imaph.tu-bs.de/qi/links.html}{M. B. Plenio}}
\qpfirstdate{18 Dec 2000}
\qplastdate{\nothing}
\qpsolve{\nothing}
%%tth: \end{html}

%
% Für Vorschau-Version nur das nötigste einstellen
%
% \nonstopmode kann ggf. auskommentiert werden!
%
\else
\nonstopmode
\documentclass[12pt,a4paper]{article}
\usepackage[latin1]{inputenc}
\markboth{\centerline{\textsl{\bfseries Draft of QI open problem -- content test only!}}}{\centerline{\textsl{\bfseries Draft of QI open problem -- content test only!}}}
\pagestyle{myheadings}
\title{Draft of QI open problem -- content test only!}
\author{No page formatting}
\fi

%
% ***************************************************************************
%
% Praeambel, redigiert durch IMaPh
%

% Keine AMS-Erweiterungen!
% Standardbefehle (\text{}, \R, \C) sind bereits in qiheader.tex
% definiert.

% TTH fuegt Label "BEGINDOCUMENT" ein und ignoriert den Block
%%tth: \iftth BEGINDOCUMENT \begin{document} \else
\ifx\quantph\undefined
	\begin{document}
\fi
\maketitle
% \makeinfo
%%tth: \fi

%
% ***************************************************************************
%
% Text-Teil, redigiert durch IMaPh
%

%
% NICHT VERGESSEN:
%	- ggf. $...$ durch \(...\) ersetzen
%	- Eintraege ins Literaturverzeichnis vollstaendig klammern: \bibitem[ABC]{4:ABC}{A.A.,B.B., ...}
%

\section{Problem}

With a Theorem by Nielsen \cite{4:N}, we have a completely explicit
criterion to decide, when one pure bipartite state can be converted to another such
state, using only local quantum operations and classical communication. Using
Nielsen's criterion one can show \cite{4:JP1} that the following strange
situation can happen: state \(A\) cannot be converted to state \(B\), but \(A\otimes
C \) can be
converted to \(B\otimes C\), where \(C\) is a suitably chosen entangled state, the
"catalyst".

The problem is to give a similarly efficient criterion to decide which pure
bipartite states can be converted into each other using a catalyst.

\section{Background}

Here is Nielsen's criterion, which is a surprisingly direct rendering of the
intuition that a ``more entangled'' pure state has a ``more mixed'' restriction.
Thus \(A\) can be converted to \(B\) if and only if the eigenvalue sequence of the
restriction of \(A\) is more mixed than that of \(B\) in the sense of majorization
of probability vectors \cite{4:Maj}. We say that one probability vector
\(p=(p_1,\ldots,p_n)\) is \emph{more mixed} than another,
\(q=(q_1,\ldots,q_n)\) in the sense of majorization, if one and
hence all of the following equivalent statements hold:

\begin{itemize}

\item For all
\(k:\quad \sum_{i>k} p_i \le \sum_{i>k}
q_i\), provided both \(p\) and \(q\) are first brought into decreasing
order.

\item
there is a doubly stochastic matrix \(D\) (positive entries, sum of all rows
and all columns = 1) such that \(p=D q\).

\item
For every convex function \(f : \R\rightarrow \R: \quad \sum_i
f(p_i) \le \sum_i f(q_i)\)

\end{itemize}

The above problem can be rephrased completely in this context of majorization
of classical probability vectors, since tensoring pure bipartite states
means again tensoring of probability vectors for the eigenvalues of the
reduced density operators. Thus we would like to characterize the order
relation ``catalytic majorization'':

\begin{quote}
For some \(r\), \((p\otimes r)\) is more mixed than \((q\otimes
r)\) in the sense of majorization.
\end{quote}

The above list of equivalent characterizations of majorization points
to a way a characterization might look like: we might look for convex functions
\(f\), such that \(p \mapsto \sum_i f(p_i)\) is monotone with respect
to catalytic majorization, and hope to characterize the relation by such
a set. One class of functions f with this monotonicity property is \(f(t)=t^x\),
for \(x>1\), because the corresponding functionals on probability vectors are
multiplicative with respect to tensor products.

There is some further literature on the use of majorization for the
characterization of pure state entanglement \cite{4:V1}, \cite{4:JP2}, \cite{4:VJN1}, 
\cite{4:N2} and on catalysis \cite{4:EW1} that may be useful.

\ifx\quantph\undefined
	\end{document}
\fi

%
% Schalter für End-Version: \let\final\relax
% Schalter für Vorschau-Version: \final nicht definiert
%
\ifx\final\relax

%
% Header-Datei laden (voller Pfad, kein Endung ".tex")
%

\ifx\quantph\undefined
\else
% Automatische Definition der Variablen
\qpvar{\qpno}
\qpvar{\qptitle}
\qpvar{\qpdate}
\qpvar{\qpcontribute}
\qpvar{\qpfirstdate}
\qpvar{\qplastdate}
\qpvar{\qpsolve}
\fi

%
% ***************************************************************************
%
% Variablenbelegungen in der LaTeX-Datei,
% einzufuegen durch IMaPh
%
% Diese Variablen sollten NICHT im TEXT-Teil
% definiert oder verwendet werden!
%

% Der folgende Block wird von TTH "verbatim" ausgegeben,
% alle anderen verarbeiten ihn
%%tth: \begin{html}
\qpno{5}
\qptitle{Maximally entangled mixed states}
\qpdate{04 Apr 2005}
\qpcontribute{\href{http://www.imaph.tu-bs.de/qi/links.html}{K. Audenaert}}
\qpfirstdate{08 Nov 2001}
\qplastdate{\nothing}
\qpsolve{\nothing}
%%tth: \end{html}

%
% Für Vorschau-Version nur das nötigste einstellen
%
% \nonstopmode kann ggf. auskommentiert werden!
%
\else
\nonstopmode
\documentclass[12pt,a4paper]{article}
\usepackage[latin1]{inputenc}
\markboth{\centerline{\textsl{\bfseries Draft of QI open problem -- content test only!}}}{\centerline{\textsl{\bfseries Draft of QI open problem -- content test only!}}}
\pagestyle{myheadings}
\title{Draft of QI open problem -- content test only!}
\author{No page formatting}
\fi

%
% ***************************************************************************
%
% Praeambel, redigiert durch IMaPh
%

% Keine AMS-Erweiterungen!
% Standardbefehle (\text{}, \R, \C) sind bereits in qiheader.tex
% definiert.

% TTH fuegt Label "BEGINDOCUMENT" ein und ignoriert den Block
%%tth: \iftth BEGINDOCUMENT \begin{document} \else
\ifx\quantph\undefined
	\begin{document}
\fi
\maketitle
% \makeinfo
%%tth: \fi

%
% ***************************************************************************
%
% Text-Teil, redigiert durch IMaPh
%

%
% NICHT VERGESSEN:
%	- ggf. $...$ durch \(...\) ersetzen
%	- Eintraege ins Literaturverzeichnis vollstaendig klammern: \bibitem[ABC]{5:ABC}{A.A.,B.B., ...}
%

\section{Problem}

Among all density operators of two qubits with the same spectrum
one may look for those maximizing some measure of entanglement.
It turns out \cite{5:VAM} that for `entanglement of formation', `relative
entropy of entanglement' and `negativity' one gets the same ``maximally
entangled states''.

Is this true for arbitrary entanglement monotones?

Obvious variants of this problem are for higher dimensional
systems and weaker constraints on the spectrum, e.\,g., largest
eigenvalue or entropy.

\section{Background}

(Refer to definitions of the measures of entanglement and `entanglement
monotone'.)

\ifx\quantph\undefined
	\end{document}
\fi

%\newpage\centerline{}

%
% Schalter für End-Version: \let\final\relax
% Schalter für Vorschau-Version: \final nicht definiert
%
\ifx\final\relax
%
% Header-Datei laden (voller Pfad, kein Endung ".tex")
%

\ifx\quantph\undefined
\else
% Automatische Definition der Variablen
\qpvar{\qpno}
\qpvar{\qptitle}
\qpvar{\qpdate}
\qpvar{\qpcontribute}
\qpvar{\qpfirstdate}
\qpvar{\qplastdate}
\qpvar{\qpsolve}
\fi

%
% ***************************************************************************
%
% Variablenbelegungen in der LaTeX-Datei,
% einzufuegen durch IMaPh
%
% Diese Variablen sollten NICHT im TEXT-Teil
% definiert oder verwendet werden!
%

% Der folgende Block wird von TTH "verbatim" ausgegeben,
% alle anderen verarbeiten ihn
%%tth: \begin{html}
\qpno{6}
\qptitle{Nice error bases}
\qpdate{04 Apr 2005}
\qpcontribute{\href{http://www.imaph.tu-bs.de/staff/dsl.html}{D. Schlingemann}}
\qpfirstdate{08 Nov 2001}
\qplastdate{28 May 2003}
\qpsolve{A.~Klappenecker, M.~Roetteler}
%%tth: \end{html}

%
% Für Vorschau-Version nur das nötigste einstellen
%
% \nonstopmode kann ggf. auskommentiert werden!
%
\else
\nonstopmode
\documentclass[12pt,a4paper]{article}
\usepackage[latin1]{inputenc}
\markboth{\centerline{\textsl{\bfseries Draft of QI open problem -- content test only!}}}{\centerline{\textsl{\bfseries Draft of QI open problem -- content test only!}}}
\pagestyle{myheadings}
\title{Draft of QI open problem -- content test only!}
\author{No page formatting}
\fi

%
% ***************************************************************************
%
% Praeambel, redigiert durch IMaPh
%

% Keine AMS-Erweiterungen!
% Standardbefehle (\text{}, \R, \C) sind bereits in qiheader.tex
% definiert.

% TTH fuegt Label "BEGINDOCUMENT" ein und ignoriert den Block
%%tth: \iftth BEGINDOCUMENT \begin{document} \else
\ifx\quantph\undefined
	\begin{document}
\fi
\maketitle
% \makeinfo
%%tth: \fi

%
% ***************************************************************************
%
% Text-Teil, redigiert durch IMaPh
%

%
% NICHT VERGESSEN:
%	- ggf. $...$ durch \(...\) ersetzen
%	- Eintraege ins Literaturverzeichnis vollstaendig klammern: \bibitem[ABC]{6:ABC}{A.A.,B.B., ...}
%

\section{Problem}

There are two special constructions to obtain \emph{orthogonal
bases of unitaries}, i.\,e., collections of unitary operators
$U_i$, $i=1,\ldots,d^2$, on a
$d$-dimensional Hilbert space, such that
$\text{tr}(U_i^* U_j) = d \delta_{i j}$:

On the one hand one can require in addition that the product of any two unitaries in the basis gives another one up to a phase, i.\,e., $U_i U_j = \text{phase} \cdot U_k$. The composition of labels $(i,i) \mapsto k$ then defines a group, the ``index group'' of the basis. Bases of this kind have been called \emph{nice error bases}.

On the other hand, one may require that, in a suitable basis of the Hilbert space, the unitaries are obtained as the products of a collection of $d$ permutation operators and $d$ multiplication operators. Bases constructed in this way are called of \emph{shift and multiply type}.

The question that arises here is to decide whether every nice error basis is of shift and multiply type. 

\section{Background}

Orthogonal bases are precisely \cite{6:3} what is needed to construct schemes for entanglement assisted teleportation or dense coding. For qubits ($d=2$) there is only one such basis up to left and right multiplication by fixed unitaries, namely the Pauli matrices together with the identity.

The shift and multiply constructions can be classified further: for the ``shift'' part one precisely needs a \emph{Latin square}, whereas the multiplication part requires the construction of $d$ complex \emph{Hadamard matrices} \cite{6:3}.

A finite group $H$ is called of \emph{central type} if it possesses an irreducible representation in $d=\sqrt{|H/Z(H)|}$ dimensions, where $Z(H)$ is the center of $H$. 

\section{Solution}

An answer to the question, given above, has recently be found by Andreas Klappenecker and Martin Roetteler. They show in their article ``On the monomiality of nice error basis'' \cite{6:4} that there is in fact a nice error basis which is not of shift and multiplier type. 

Roughly their argumentation is based on the following: First one observes that
every nice error basis which is of shift and multiplier type is monomial, i.e.
each of its unitary matices has in every row and column precisely one
non-vanishing entry. An abstract error group is one which is generated by nice
error bases (central extension of the index group). Such a group is of
\emph{central type} with \emph{cyclic center}. Employing the theory of
characters for these groups, which has been studied by P.~Ferguson, I.\,M.~Isaacs (see references given in \cite{6:4}), an abstract error group can be constructed which has a non-monomial irreducible representation.

\ifx\quantph\undefined
	\end{document}
\fi

%
% Schalter für End-Version: \let\final\relax
% Schalter für Vorschau-Version: \final nicht definiert
%
\ifx\final\relax

%
% Header-Datei laden (voller Pfad, kein Endung ".tex")
%

\ifx\quantph\undefined
\else
% Automatische Definition der Variablen
\qpvar{\qpno}
\qpvar{\qptitle}
\qpvar{\qpdate}
\qpvar{\qpcontribute}
\qpvar{\qpfirstdate}
\qpvar{\qplastdate}
\qpvar{\qpsolve}
\fi

%
% ***************************************************************************
%
% Variablenbelegungen in der LaTeX-Datei,
% einzufuegen durch IMaPh
%
% Diese Variablen sollten NICHT im TEXT-Teil
% definiert oder verwendet werden!
%

% Der folgende Block wird von TTH "verbatim" ausgegeben,
% alle anderen verarbeiten ihn
%%tth: \begin{html}
\qpno{7}
\qptitle{Additivity of Entanglement of Formation}
\qpdate{04 Apr 2005}
\qpcontribute{\href{http://www.imaph.tu-bs.de/qi/links.html}{K. G. H. Vollbrecht}}
\qpfirstdate{16 Nov 2001}
\qplastdate{11 Nov 2004}
\qpsolve{(equivalent to \protect\hyperlink{Outline10.0}{problem 10})}
%%tth: \end{html}

%
% Für Vorschau-Version nur das nötigste einstellen
%
% \nonstopmode kann ggf. auskommentiert werden!
%
\else
\nonstopmode
\documentclass[12pt,a4paper]{article}
\usepackage[latin1]{inputenc}
\markboth{\centerline{\textsl{\bfseries Draft of QI open problem -- content test only!}}}{\centerline{\textsl{\bfseries Draft of QI open problem -- content test only!}}}
\pagestyle{myheadings}
\title{Draft of QI open problem -- content test only!}
\author{No page formatting}
\fi

%
% ***************************************************************************
%
% Praeambel, redigiert durch IMaPh
%

% Keine AMS-Erweiterungen!
% Standardbefehle (\text{}, \R, \C) sind bereits in qiheader.tex
% definiert.

% TTH fuegt Label "BEGINDOCUMENT" ein und ignoriert den Block
%%tth: \iftth BEGINDOCUMENT \begin{document} \else
\ifx\quantph\undefined
	\begin{document}
\fi
\maketitle
% \makeinfo
%%tth: \fi

%
% ***************************************************************************
%
% Text-Teil, redigiert durch IMaPh
%

%
% NICHT VERGESSEN:
%	- ggf. $...$ durch \(...\) ersetzen
%	- Eintraege ins Literaturverzeichnis vollstaendig klammern: \bibitem[ABC]{7:ABC}{A.A.,B.B., ...}
%
\section{Problem}

The \emph{entanglement of formation} \cite{7:BD96} is one  of the
standard measures of entanglement. It is defined, for any density operator
\(\rho\) on a bipartite system, as
\[    E_\text{\scriptsize F}(\rho)=
       \text{inf} \left\{ \sum_i
           r_i S(\rho_i|A)
        \,\Big|\, \sum_i
           r_i \rho_i = \rho \right\},
\]
where \(S(.)\) denotes the von Neumann entropy and \(\rho|A\) denotes the
restriction of a density operator \(\rho\) to the ``Alice''
subsystem (partial trace over the other subsystem), the
\(\rho_i\) are density operators and the
\(r_i\) are positive, adding up to one. Since \(S\) is
concave, the infimum is attained at a convex decomposition of \(\rho\)
into pure states, and the definition is often given as this
restricted infimum.

Consider now a pair \(\rho^{(i)}\), \(i=1,2\) of
bipartite density operators, and their tensor product
\(\rho=\rho^{(1)}\otimes\rho^{(2)}\),
which lives on a tensor product of four Hilbert spaces, but can be
considered as a bipartite state when the two Alice subsepaces and
the two Bob subspaces are grouped together. Then it is easy to
show (by plugging the tensor product of the optimal decompositions
of the factors into the variational expression and using the
additivity of the entropy) that
\(E_\text{\scriptsize F}(\rho)
	\le E_\text{\scriptsize F}(\rho^{(1)})
       + E_\text{\scriptsize F}(\rho^{(2)})\).

The problem is to show that equality always holds here.
\section{Background}

This inequality is crucial to settle the interpretation of
\(E_\text{\scriptsize F}\) as a \emph{resource} quantity. The typical kind of
tensor products appearing in the theory are pairs created by
(maybe different) sources of entangled states, and kept for later
use.

\section{Partial Solutions}

The additivity of entanglement of formation could be proven for several
examples of states by Vidal et al. \cite{7:VDC02}.

This problem has been shown to be equivalent to \hyperlink{Outline10.0}{problem 10}.

\ifx\quantph\undefined
	\end{document}
\fi

%
% Schalter für End-Version: \let\final\relax
% Schalter für Vorschau-Version: \final nicht definiert
%
\ifx\final\relax

%
% Header-Datei laden (voller Pfad, kein Endung ".tex")
%

\ifx\quantph\undefined
\else
% Automatische Definition der Variablen
\qpvar{\qpno}
\qpvar{\qptitle}
\qpvar{\qpdate}
\qpvar{\qpcontribute}
\qpvar{\qpfirstdate}
\qpvar{\qplastdate}
\qpvar{\qpsolve}
\fi

%
% ***************************************************************************
%
% Variablenbelegungen in der LaTeX-Datei,
% einzufuegen durch IMaPh
%
% Diese Variablen sollten NICHT im TEXT-Teil
% definiert oder verwendet werden!
%

% Der folgende Block wird von TTH "verbatim" ausgegeben,
% alle anderen verarbeiten ihn
%%tth: \begin{html}
\qpno{8}
\qptitle{Qubit formula for Relative Entropy of Entanglement}
\qpdate{04 Apr 2005}
\qpcontribute{\href{http://www.imaph.tu-bs.de/qi/links.html}{J. Eisert}}
\qpfirstdate{20 Jun 2003}
\qplastdate{\nothing}
\qpsolve{\nothing}
%%tth: \end{html}

%
% Für Vorschau-Version nur das nötigste einstellen
%
% \nonstopmode kann ggf. auskommentiert werden!
%
\else
\nonstopmode
\documentclass[12pt,a4paper]{article}
\usepackage[latin1]{inputenc}
\markboth{\centerline{\textsl{\bfseries Draft of QI open problem -- content test only!}}}{\centerline{\textsl{\bfseries Draft of QI open problem -- content test only!}}}
\pagestyle{myheadings}
\title{Draft of QI open problem -- content test only!}
\author{No page formatting}
\fi

%
% ***************************************************************************
%
% Praeambel, redigiert durch IMaPh
%

% Keine AMS-Erweiterungen!
% Standardbefehle (\text{}, \R, \C) sind bereits in qiheader.tex
% definiert.

% TTH fuegt Label "BEGINDOCUMENT" ein und ignoriert den Block
%%tth: \iftth BEGINDOCUMENT \begin{document} \else
\ifx\quantph\undefined
	\begin{document}
\fi
\maketitle
% \makeinfo
%%tth: \fi

%
% ***************************************************************************
%
% Text-Teil, redigiert durch IMaPh
%

%-ltoh-	:comm	:\inf:inf::

%
% NICHT VERGESSEN:
%	- ggf. $...$ durch \(...\) ersetzen
%	- Eintraege ins Literaturverzeichnis vollstaendig klammern: \bibitem[ABC]{8:ABC}{A.A.,B.B., ...}
%

\section{Problem}
The \emph{relative entropy of entanglement} is an entanglement monotone that
quantifies to what extent a given state can be operationally distinguished (in the
sense of Stein's Lemma) from the closest state which is either separable or has a
positive partial transpose (PPT). For a state $\rho$ it is defined as \cite{8:1}
\[ E_R(\rho) = \inf_{\sigma \in D} S(\rho||\sigma), \]
where  $D$ stands for the convex sets of separable or PPT states, and $S(.||.)$ is
the quantum relative entropy. The problem is to find a closed formula for this
quantity for systems consisting of two qubits.

\section{Background}
The interpretation of the relative entropy of entanglement is a geometrical one:
it is related to the error probability with which a state is mistakenly assumed to
be merely classically correlated or PPT in quantum hypothesis testing. This
entanglement monotone is an upper bound to the distillable entanglement, and in
its asymptotic version conjectured to be identical to the Rains' bound for
distillable entanglement. As most other monotones of entanglement, and all other
known monotones that are provably asymptotically continuous, the actual evaluation
of this quantity amounts to solving an optimization problem. In the case at hand,
it is a convex optimization problem.

The \emph{entanglement of formation} is a monotone which is also defined as an
optimization problem. If it turned out that the entanglement of formation was in
fact additive (see problem \href{7.html}{7}), then this quantity could be
interpreted as the \emph{entanglement cost}, which fleshes out  the resource
character of entanglement. Historically, it was very important that for systems
consisting of two qubits, the entanglement of formation can (quite astonishingly)
be evaluated: the Wootters formula \cite{8:2} is a closed formula for the
entanglement of formation for two-qubit systems. The proof exploits a number of
the particular properties that are available for two-qubit systems \cite{8:3} -- and
only for them. The task is to explicitly solve the convex optimization problem
posed by the \emph{relative entropy of entanglement}.

\section{Partial Solution}
So far, there is no published solution to the problem. Ref. \cite{8:4} presents the
solution to a related problem: for a two-qubit system, given a state on the
boundary of separable states  $\sigma$, it characterizes the states $\rho$ for
which $E_R(\rho)=S(\rho||\sigma)$.

\ifx\quantph\undefined
	\end{document}
\fi

%
% Schalter für End-Version: \let\final\relax
% Schalter für Vorschau-Version: \final nicht definiert
%
\ifx\final\relax

%
% Header-Datei laden (voller Pfad, kein Endung ".tex")
%

\ifx\quantph\undefined
\else
% Automatische Definition der Variablen
\qpvar{\qpno}
\qpvar{\qptitle}
\qpvar{\qpdate}
\qpvar{\qpcontribute}
\qpvar{\qpfirstdate}
\qpvar{\qplastdate}
\qpvar{\qpsolve}
\fi

%
% ***************************************************************************
%
% Variablenbelegungen in der LaTeX-Datei,
% einzufuegen durch IMaPh
%
% Diese Variablen sollten NICHT im TEXT-Teil
% definiert oder verwendet werden!
%

% Der folgende Block wird von TTH "verbatim" ausgegeben,
% alle anderen verarbeiten ihn
%%tth: \begin{html}
\qpno{9}
\qptitle{Reduction criterion implies majorization?}
\qpdate{04 Apr 2005}
\qpcontribute{\href{http://www.imaph.tu-bs.de/qi/links.html}{M. M. Wolf}}
\qpfirstdate{12 Feb 2002}
\qplastdate{20 May 2003}
\qpsolve{T. Hiroshima}
%%tth: \end{html}

%
% Für Vorschau-Version nur das nötigste einstellen
%
% \nonstopmode kann ggf. auskommentiert werden!
%
\else
\nonstopmode
\documentclass[12pt,a4paper]{article}
\usepackage[latin1]{inputenc}
\markboth{\centerline{\textsl{\bfseries Draft of QI open problem -- content test only!}}}{\centerline{\textsl{\bfseries Draft of QI open problem -- content test only!}}}
\pagestyle{myheadings}
\title{Draft of QI open problem -- content test only!}
\author{No page formatting}
\fi

%
% ***************************************************************************
%
% Praeambel, redigiert durch IMaPh
%

% Keine AMS-Erweiterungen!
% Standardbefehle (\text{}, \R, \C) sind bereits in qiheader.tex
% definiert.
\iftth\else\newcommand{\majorize}{\prec}\fi

% TTH fuegt Label "BEGINDOCUMENT" ein und ignoriert den Block
%%tth: \iftth BEGINDOCUMENT \begin{document} \else
\ifx\quantph\undefined
	\begin{document}
\fi
\maketitle
% \makeinfo
%%tth: \fi

%
% ***************************************************************************
%
% Text-Teil, redigiert durch IMaPh
%

%
% NICHT VERGESSEN:
%	- ggf. $...$ durch \(...\) ersetzen
%	- Eintraege ins Literaturverzeichnis vollstaendig klammern: \bibitem[ABC]{9:ABC}{A.A.,B.B., ...}
%

\section{Problem}

The density matrix of any separable state is majorized by its reductions (the
density matrix reduced to one subsystem, e.\,g. ${{\rho}_{A}} = Tr_{B}\,
{{\rho}_{AB}}$). This is in fact the strongest separability criterion based on
the spectra of a state and one of its reductions. However, it is not known how
it is related to other separability criteria like PPT, undistillability or the
reduction criterion. The problem is to find out how majorization enters into
the known implication chain of separability criteria.

\section{Background}

One of the remarkable properties of entangled states is that they
can exhibit locally more disorder than globally. The simplest
example is the maximally entangled state, which is pure as a whole
but it has maximally chaotic reductions. A powerful tool comparing
the order/disorder of two systems is majorization and in fact it
is a more stringent notion of order/disorder than entropy.

It was proven in \cite{9:1} that the density matrix of a separable
state is majorized by both of its reductions. Hence, majorization yields a
separability criterion, which is merely based on the spectra of a state and its
reductions.

There are many important separability/entanglement criteria or properties and
in most cases the relations between them are well known: Separability \imply\ 
positivity of the partial transpose \cite{9:2} \imply\ undistillability \cite{9:3}
\imply\ reduction criterion \cite{9:4}.

The intuition may be, that all these criteria a strictly stronger
than majorization, however the matter is not decided yet.

\section{Partial Solutions}

Apart from inconclusive numerical search for counterexamples for the
implication: reduction criterion \imply\ majorization, the only partial result
is derived in \cite{9:5}, where it was shown, that the reduction criterion
implies positivity for conditional Renyi entropies for every value of the
entropic parameter. Although conditional entropies also measure the proportion
between global and local disorder, this result cannot be extended directly to
majorization.

\section{Solution}

The answer to the question is contained in \cite{9:6}, stating that the reduction criterion \emph{does imply} majorization.

The key idea of the proof is that ${{\rho}_{A}} \otimes I_{B} \geq {{\rho}_{AB}}$
implies ${\rho}_{AB}^{1/2} = ({\rho}_{A}^{1/2} \otimes I_{B})\, R$ with $\left\|
R\right\| \leq 1$, where ${{\rho}_{AB}}$ is a bipartite density matrix,
${{\rho}_{A}} = Tr_{B}\, {{\rho}_{AB}}$, and $\left\| \cdot \right\|$ is the
operator norm. By virtue of this, we can derive the existence of the substochastic
matrix $S$ such that $\lambda({{\rho}_{AB}}) = S\, \lambda({{\rho}_{A}})$, where
$\lambda({{\rho}_{AB}})$ [$\lambda({{\rho}_{A}})$] is the eigenvalue (column)
vector of ${{\rho}_{AB}}$ [${{\rho}_{A}}$]. This last equation is equivalent to
the weak submajorization relation $\lambda({{\rho}_{AB}}) {\majorize}_{w}
\lambda({{\rho}_{A}})$ which is none other than $\lambda({{\rho}_{AB}}) \majorize
\lambda({{\rho}_{A}})$ in this problem.

\ifx\quantph\undefined
	\end{document}
\fi

%
% Schalter fuer End-Version: \let\final\relax
% Schalter fuer Vorschau-Version: \final nicht definiert
%
\ifx\final\relax

%
% Header-Datei laden (voller Pfad, kein Endung ".tex")
%

\ifx\quantph\undefined
\else
% Automatische Definition der Variablen
\qpvar{\qpno}
\qpvar{\qptitle}
\qpvar{\qpdate}
\qpvar{\qpcontribute}
\qpvar{\qpfirstdate}
\qpvar{\qplastdate}
\qpvar{\qpsolve}
\fi

%
% ***************************************************************************
%
% Variablenbelegungen in der LaTeX-Datei,
% einzufuegen durch IMaPh
%
% Diese Variablen sollten NICHT im TEXT-Teil
% definiert oder verwendet werden!
%

% Der folgende Block wird von TTH "verbatim" ausgegeben,
% alle anderen verarbeiten ihn
%%tth: \begin{html}
\qpno{10}
\qptitle{Additivity of classical capacity and related problems}
\qpdate{04 Apr 2005}
\qpcontribute{\href{http://www.imaph.tu-bs.de/qi/links.html}{A. S. Holevo}}
\qpfirstdate{31 Jan 2003}
\qplastdate{11 Nov 2004}
\qpsolve{\nothing}
%%tth: \end{html}

%
% Fuer Vorschau-Version nur das noetigste einstellen
%
% \nonstopmode kann ggf. auskommentiert werden!
%
\else
\nonstopmode
\documentclass[12pt,a4paper]{article}
\usepackage[latin1]{inputenc}
\markboth{\centerline{\textsl{\bfseries Draft of QI open problem -- content test only!}}}{\centerline{\textsl{\bfseries Draft of QI open problem -- content test only!}}}
\pagestyle{myheadings}
\title{Draft of QI open problem -- content test only!}
\author{No page formatting}
\fi

%
% ***************************************************************************
%
% Praeambel, redigiert durch IMaPh
%

% Keine AMS-Erweiterungen!
% Standardbefehle (\text{}, \R, \C) sind bereits in qiheader.tex
% definiert.

% Fuer Tabelle
% \newcommand{\tpbox}[2]{\parbox{#1}{\centering
%   \pretolerance=-1\tolerance=0\hyphenpenalty=0\doublehyphendemerits=0
%   \finalhyphendemerits=0\exhyphenpenalty=0 #2}}
% \newcommand{\nhyp}{\hspace{0pt plus0pt minus0pt}}
% %-ltoh-   :comm   :\nhyp:::
% %-ltoh-   :{2}    :\tpbox:#2::
%%tth:  \iftth
%%tth:  \newcommand{\fparbox}[2]{\special{html:<table align="center" border="1" cellpadding="4"><tr><td>}#2\special{html:</td></tr></table>}}
%%tth:  \newcommand{\cl}[1]{\special{html:<center>}#1\special{html:</center>}}
%%tth:  \else
\providecommand{\fparbox}[2]{\fbox{\parbox{#1}{#2}}}
\providecommand{\cl}[1]{\centerline{#1}}
%-ltoh- :{2}    :\fparbox:<table align="center" border="1" cellpadding="4"><tr><td>#2</td></tr></table>::
%-ltoh- :{}     :\cl:<center>:</center>:
%%tth:  \fi

%-ltoh- :b/e    :\begin{equation}:<p align=center>:</p>:

% TTH fuegt Label "BEGINDOCUMENT" ein und ignoriert den Block
%%tth: \iftth BEGINDOCUMENT \begin{document} \else
\ifx\quantph\undefined
	\begin{document}
\fi
\maketitle
% \makeinfo
%%tth: \fi

%
% ***************************************************************************
%
% Text-Teil, redigiert durch IMaPh
%

%
% NICHT VERGESSEN:
%   - ggf. $...$ durch \(...\) ersetzen
%   - Eintraege ins Literaturverzeichnis vollstaendig klammern: \bibitem[ABC]{10:ABC}{A.A.,B.B., ...}
%

\section{Problem}

For each quantum channel $T$ (in the Schr\"odinger picture), define

\[
\chi(T) = \text{sup}_{p,\rho}\left( H\left(\sum_i p_i
T(\rho_i)\right) -\sum_i p_i H\left(T(\rho_i)\right) \right),
\]

where the supremum is over all probability vectors $p=(p_1,\ldots,p_n)$, and
all collections of input states $\{\rho_1,\ldots,\rho_n\}$, and $H$ denotes
the von Neumann entropy.

Show that $\chi (T_1\otimes T_2)=\chi (T_1)+\chi (T_2)$, or else give a
counterexample. The problem can be traced back to \cite{10:BFS}, see also
\cite{10:Ho}.

\section{Background}

This problem can also be paraphrased as ``Can entanglement between signal states
help to send classical information on quantum channels?''.

Recall that the capacities of a memoryless channel are defined as the maximal
transmission rate per use of the channel, with coding and decoding
chosen for increasing number $n$ of parallel and
independent uses of the channel
\[
T^{\otimes n}=\underset{n}{\underbrace{T\otimes \dots \otimes T}}
\]
such that the error probability goes to zero as $n\rightarrow \infty
$. There are many different capacities, because one may consider
sending different kinds (classical or quantum) information, restrict
the admissible coding and decoding operations, and/or allow the use
of additional resources. Here we only look at the transmission of
classical information with no additional resources. Then one can
distinguish four capacities \cite{10:BS}, according to whether for
each block length $n$ we are allowed to use arbitrary entangled
quantum operations on the full block of input (resp. output)
systems, or if for each of the parallel channels we have to use a
separate quantum coding (resp. decoding), and combine these only by
classical pre (resp. post)-processing:

\begin{center}
{\scriptsize
\begin{tabular}{ccccc}
&  & \fbox{
\parbox{3.7cm}{\cl{$C_{\infty \infty}$: \emph{full
capacity},}\cl{arbitrary (de)coding}}} &  &  \\
& ??? &  & $\ge$ &  \\
\fbox{
\parbox{3.0cm}{
\cl{$C_{1 \infty} = \chi$:} \cl{unentangled coding,} \cl{quantum
block decoding} }} &  &  &  & \fbox{
\parbox{3.0cm}{
\cl{$C_{\infty 1}$:} \cl{quantum block coding,} \cl{separate
decoding}
}} \\
& $\ge$ &  & $=$ &  \\
&  & \fbox{
\parbox{3.7cm}{
\cl{$C_{1 1}$: \emph{one-shot capacity}} \cl{or \emph{accessible
information},} \cl{separate quantum (de)coding,} \cl{block
(de)coding only classical} }} &  &
\end{tabular}
}
\end{center}

The equality in the lower right was established independently by
several authors, see e. g. \cite{10:KRb}. That $C_{1\infty }$ on the
left coincides with the quantity $\chi $ given in the statement of
the problem was shown in \cite{10:HSW}. The inequality in the lower
left is known to be strict sometimes \cite{10:Ho}, which means that
entangling decodings indeed can increase the classical capacity. See
\cite{10:SKIH} for investigation of the corresponding information gain.
The full capacity and $\chi $ are connected by the limit formula

\[
C_{\infty \infty}(T) = \mathrm{lim} _n (1/n) \chi(T^{\otimes n})
\]

Since $\chi $ is easily seen to be superadditive (i.\, e., $\chi (T_{1}\otimes
T_{2})\geq \chi (T_{1})+\chi (T_{2})$), we immediately get $ C_{\infty \infty
}\geq \chi $. If additivity holds, then we will even have equality, i.\, e.,
``???'' in the table can be replaced by ``$=$'' . While such a result would be
very much welcome from a mathematical (and practical) point of view, giving a
``single-letter'' expression for the classical capacity, it would call for a
physical explanation of strange asymmetry between the roles of entanglement in
encoding and decoding procedures.

\section{Partial results}

Validity of the additivity conjecture was established if one of the
channels is

\begin{itemize}
\item the identity channel \cite{10:AHW}, \cite{10:SWa};

\item a unital qubit channel \cite{10:Kib};

\item the depolarizing channel \cite{10:Kic};

\item an entaglement breaking channel \cite{10:Ho}, \cite{10:Kia} (both for ``c-q/q-c''
	channels), \cite{10:Sha} (general entaglement breaking channel).
\end{itemize}

Some further more recent partial results will be mentioned below. Whether the
additivity holds ``globally'', i.\,e. for all quantum channels, is still an open
problem. No counterexample was found despite extensive numerical search by groups
in IBM, IMaPh, see also \cite{10:ON}. If the conjecture is valid, then the
additivity of $\chi $ tentatively relies upon yet another hypothetical property of
multiplicativity of norms of the completely positive mappings
\[
T:\ell _{1}(\mathcal{H})\rightarrow \ell _{p}(\mathcal{H});\quad
p\geq 1,
\]
where
\[
\ell _{p}(\mathcal{H})=\{X:X=X^{\ast },\quad \Vert X\Vert _{p}\equiv
\left( \mathrm{Tr}|X|^{p}\right) ^{\frac{1}{p}}\}
\]
is a noncommutative analog of the space $\ell _{p}$ -- the so called
Schatten class. Namely, the conjecture \cite{10:AHW} is that for $p$,
sufficiently close to $1$
\begin{equation}
\Vert T_{1}\otimes T_{2}\Vert _{p}\overset{?}{=}\Vert T_{1}\Vert
_{p}\Vert T_{2}\Vert _{p},  \label{10:mult}
\end{equation}
where $\Vert T\Vert _{p}=\max_{\rho }\Vert T(\rho )\Vert _{p}.$ By
letting $ p\downarrow 1$ this implies additivity of the minimal
output entropy
\[
H_{\min }(T)=\min_{\rho }H\left( T\left( \rho \right) \right) ,
\]
one of a whole number of properties equivalent, as it was shown in
\cite{10:Shb} , to the additivity of $\chi $. The relation (\ref{10:mult})
can be re-expressed as the additivity of the minimal output Renyi
entropy of order $ p$ \cite{10:GGLMSY}.

In all cases listed above where the additivity conjecture is proved,
the multiplicativity of $p-$norms (for all $p\geq 1$) also holds,
moreover, it underlies the proof of additivity in \cite{10:Kib},
\cite{10:Kic}. The multiplicativity of $p-$norms holds for arbitrary
bounded maps of the classical spaces $\ell _{p}$, where its proof
can be based on a Minkowsky inequality. Therefore quite intriguing
is counterexample of the channel
\[
T(\rho )=\frac{1}{d-1}\left[ I-\rho ^{T}\right] ,
\]
for which (\ref{10:mult}) with $T_{1}=T_{2}=T$ fails to hold for
sufficiently large $p$ ($p\geq 4,7823$ if
$d=\mathrm{dim}\mathcal{H}=3$ \cite{10:WH}). Nevertheless, the
additivity of $H_{\min }$ and of $\chi $ holds for such channels, as
shown in \cite{10:MY}, \cite{10:DHS}, \cite{10:AF}. The standing conjecture
is that multiplicativity holds globally at least for $1\leq p\leq
2,$ but  even the case $p=2$ is difficult, see \cite{10:KNR},
\cite{10:KRc}. For some results concerning integer $p$ see \cite{10:GLR}.

In \cite{10:AB} it was shown that proving the multiplicativity would
solve another important open problem -- superadditivity of the
entanglement of formation (EoF). Earlier \cite{10:MSW} brought
attention to a simple correspondence between $\chi $ and EoF, and
obtained several concrete results on additivity of EoF by using this
correspondence. It was also remarked that superadditivity of EoF
would imply additivity of $\chi $ for channels with linear additive
input constraints. By combining the MSW correspondence and the
convex duality technique of \cite{10:AB} with an original and powerful
channel extension technique, which allows to use effectively
arbirariness of channels in question, \cite{10:Shb} had shown
equivalence of the global properties of additivity of the minimal
output entropy, $\chi ,$ EoF and of superadditivity of EoF. The last
equivalence for two fixed channels was also established in
\cite{10:Po}.

In \cite{10:HSa} several equivalent formulations of the additivity
conjecture for  channels with arbitrarily constrained inputs, which
formally is substantially stronger than  additivity of the
unconstrained $\chi $, were given. It was shown that the additivity
conjecture for channels with constrained inputs holds true for
certain nontrivial classes of channels, e. g. a direct sum mixture
of the identity channel and an entaglement breaking channel (such as
erasure channel). The channel extension technique was used to show
that additivity for two fixed constrained channels can be reduced to
the same problem for unconstrained channels, and hence, the global
additivity for channels with arbitrary input constraints is
equivalent to the global additivity without constraints.

%%tth:  \iftth
%%tth:  \special{html:<img src="fig10-1.png"><br><font size=-1>}
%%tth:	\special{html:Fig.1: Equivalence of additivity properties. Bold (thin) arrows indicate nontrivial (obvious) implications for individual (ind.) or global (glob.) relations.}
%%tth:	\special{html:</font>}
%%tth:  \else
%-ltoh- +{2}+\begin#1#2+<img src="fig10-1.png"><br><font size=-1>Fig.1: Equivalence of additivity properties. Bold (thin) arrows indicate nontrivial (obvious) implications for individual (ind.) or global (glob.) relations.</font>++
%-ltoh- :comm:\end{figure}:::

% Here starts picture of additivity relations
\begin{figure}{
    \begin{center}
    \setlength{\unitlength}{\baselineskip}
    \begin{picture}(17.5,22)
        \put(0,18){\framebox(7,3){\parbox{6\unitlength}{\begin{center}
                Superadditivity\\ of EoF
            \end{center}}}}
        \put(3,18){\vector(0,-1){3}}
        \put(0.5,16){\parbox[b]{2\unitlength}{\raggedleft ind.}}
        \thicklines
        \put(4,15){\vector(0,1){3}}
        \thinlines
        \put(4.5,16){\parbox[b]{5\unitlength}{ind. \cite{10:Shb,10:Po}}}

        \put(0,12){\framebox(7,3){\parbox{6\unitlength}{\begin{center}
                Additivity\\ of EoF
            \end{center}}}}
        \thicklines
        \put(3.5,9){\vector(0,1){3}}
        \thinlines
        \put(-1,10){\parbox[b]{4\unitlength}{\centerline{glob. \cite{10:Shb}}}}

        \put(0,6){\framebox(7,3){\parbox{6\unitlength}{\begin{center}
                Additivity\\ of $\chi$
            \end{center}}}}
        \thicklines
        \put(3.5,6){\vector(0,-1){3}}
        \thinlines
        \put(4,4){\parbox[b]{4\unitlength}{glob. \cite{10:Shb}}}

        \put(0,0){\framebox(7,3){\parbox{6\unitlength}{\begin{center}
                Additivity\\ of $H_\text{min}$
            \end{center}}}}

        \put(12,18){\framebox(7,3){\parbox{6\unitlength}{\begin{center}
                Additivity of\\ constrained $\chi$
            \end{center}}}}
        \thicklines
        \put(9.5,20){\vector(-1,0){2.5}}
        \put(9.5,20){\vector(1,0){2.5}}
        \thinlines
        \put(7,20.5){\parbox[b]{5\unitlength}{\centerline{ind. \cite{10:HSa}}}}

        \thicklines
        \put(-1.5,14){\vector(1,0){1.5}}
        \put(-1.5,2){\line(0,1){12}}
        \put(-1.5,2){\vector(1,0){1.5}}
        \thinlines

        \thicklines
        \put(7,19){\line(1,0){4}}
        \put(11,19){\line(0,-1){11}}
        \put(11,8){\vector(-1,0){4}}
        \thinlines
        \put(5.5,10){\parbox[b]{5\unitlength}{\raggedleft ind. \cite{10:MSW}}}

        \thicklines
        \put(0,20){\line(-1,0){2.5}}
        \put(-2.5,20){\line(0,-1){19}}
        \put(-2.5,1){\vector(1,0){2.5}}
        \thinlines
        \put(-2,14){\rotatebox{90}{\parbox[b]{6\unitlength}{\centerline{ind. \cite{10:Shb}}}}}

        \put(15,18){\line(0,-1){10}}
        \put(15,8){\line(-1,0){4}}
        \put(9.5,13.5){\parbox[b]{5\unitlength}{\raggedleft ind.}}
        \thicklines
        \put(16,7){\vector(0,1){11}}
        \put(7,7){\line(1,0){9}}
        \thinlines
        \put(16.5,13.5){\parbox[b]{4\unitlength}{glob. \cite{10:HSa}}}

    \end{picture}
    \end{center}
    \caption{Equivalence of additivity properties. Bold (thin) arrows indicate
    nontrivial (obvious) implications for individual
    (ind.) or global (glob.) relations.}
}\end{figure}
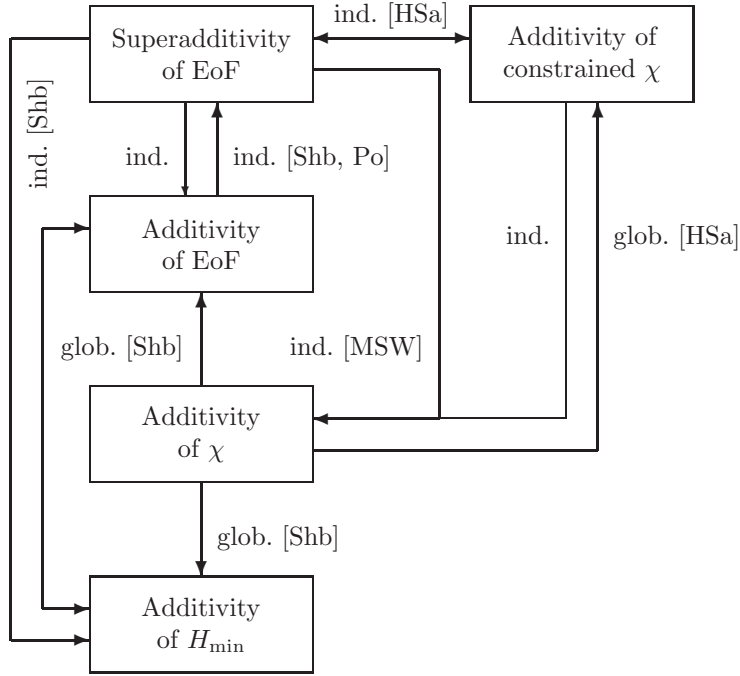
% End of picture

%%tth:  \fi

The additivity problem is still open for the minimal dimension 2: it
is not known if the additivity holds for all nonunital qubit
channels, although a strong numerical evidence in favour of this was
given in \cite{10:HIMRS}. Nevertheless there are several reasons
to consider the problem in infinite dimensions. There is a good
chance that both the additivity and the multiplicativity for all
$p\geq 1$ hold for important and interesting class of Gaussian
channels that act in infinite dimensional Hilbert space. However the
only instance where the additivity of $\chi $ and the
multiplicativity for integer $p$\ was proved is the pure loss
channel, having the very special property $H_{\min }(T)=0$
\cite{10:GGLMSY}, \cite{10:GL}$.$

It was observed recently that Shor's proof of equivalence of
different forms of the global additivity conjecture for finite
dimensional channels is related to weird discontinuity of the $\chi
-$capacity as a function of channel in infinite dimensions. This
also calls for a mathematically rigorous treatment of the entropic
quantities related to the classical capacity of infinite dimensional
channels \cite{10:HSb}.  In particular it is possible to show that
additivity for all finite dimensional channels implies additivity of
the constrained $\chi -$capacity with constraints fulfilling
finiteness of the output entropy \cite{10:Shi}.

\ifx\quantph\undefined
	\end{document}
\fi

%\newpage\centerline{}

%
% Schalter für End-Version: \let\final\relax
% Schalter für Vorschau-Version: \final nicht definiert
%
\ifx\final\relax

%
% Header-Datei laden (voller Pfad, kein Endung ".tex")
%

\ifx\quantph\undefined
\else
% Automatische Definition der Variablen
\qpvar{\qpno}
\qpvar{\qptitle}
\qpvar{\qpdate}
\qpvar{\qpcontribute}
\qpvar{\qpfirstdate}
\qpvar{\qplastdate}
\qpvar{\qpsolve}
\fi

%
% ***************************************************************************
%
% Variablenbelegungen in der LaTeX-Datei,
% einzufuegen durch IMaPh
%
% Diese Variablen sollten NICHT im TEXT-Teil
% definiert oder verwendet werden!
%

% Der folgende Block wird von TTH "verbatim" ausgegeben,
% alle anderen verarbeiten ihn
%%tth: \begin{html}
\qpno{11}
\qptitle{Continuity of the quantum channel capacity}
\qpdate{04 Apr 2005}
\qpcontribute{\href{http://www.imaph.tu-bs.de/staff/mk.html}{M. Keyl}}
\qpfirstdate{20 Jun 2003}
\qplastdate{\nothing}
\qpsolve{\nothing}
%%tth: \end{html}

%
% Für Vorschau-Version nur das nötigste einstellen
%
% \nonstopmode kann ggf. auskommentiert werden!
%
\else
\nonstopmode
\documentclass[12pt,a4paper]{article}
\usepackage[latin1]{inputenc}
\markboth{\centerline{\textsl{\bfseries Draft of QI open problem -- content test only!}}}{\centerline{\textsl{\bfseries Draft of QI open problem -- content test only!}}}
\pagestyle{myheadings}
\title{Draft of QI open problem -- content test only!}
\author{No page formatting}
\fi

%
% ***************************************************************************
%
% Praeambel, redigiert durch IMaPh
%

% Keine AMS-Erweiterungen!
% Standardbefehle (\text{}, \R, \C) sind bereits in qiheader.tex
% definiert.

% TTH fuegt Label "BEGINDOCUMENT" ein und ignoriert den Block
%%tth: \iftth BEGINDOCUMENT \begin{document} \else
\ifx\quantph\undefined
	\begin{document}
\fi
\maketitle
% \makeinfo
%%tth: \fi

%
% ***************************************************************************
%
% Text-Teil, redigiert durch IMaPh
%

%
% NICHT VERGESSEN:
%	- ggf. $...$ durch \(...\) ersetzen
%	- Eintraege ins Literaturverzeichnis vollstaendig klammern: \bibitem[ABC]{11:ABC}{A.A.,B.B., ...}
%

\section{Problem}

The quantum capacity of a noisy quantum channel can be regarded as a function 
on the space of all channels. Is this function continuous? In other words: If 
the distance (e.g. with respect to the cb-norm) between two channels is 
small, is the distance between the corresponding capacities small as well?

\section{Partial Solutions}

In \cite{11:1} it was shown that the quantum capacity as a function of the channel is
lower semi-continuous.

\ifx\quantph\undefined
	\end{document}
\fi

\addtocontents{toc}{\protect\end{tabular}}
%\addtocontents{toc}{\protect\enlargethispage{1ex}}
\addtocontents{toc}{\protect\clearpage\protect\noindent}
\addtocontents{toc}{\protect\begin{tabular}{@{}c@{}c@{}c@{}c@{}>{\protect\centering}p{25\cu}@{}c@{}}}
\addtocontents{toc}{\protect\tocback{No}&\protect\multicolumn{4}{l}{\tocback{Title}}&\protect\tocback{Page}\\%
	&\protect\tocback{Contact}&\protect\tocback{Date}&\protect\tocback{Last progress}&\protect\tocback{Solved by}%
	\protect\tabularnewline[2ex]}
\addtocontents{toc}{\hspace*{5\cu}&\hspace*{27\cu}&\hspace*{18\cu}&\hspace*{18\cu}&\hspace*{27\cu}&\hspace*{5\cu}\\}
%\newpage\centerline{}

%
% Schalter für End-Version: \let\final\relax
% Schalter für Vorschau-Version: \final nicht definiert
%
\ifx\final\relax

%
% Header-Datei laden (voller Pfad, kein Endung ".tex")
%

\ifx\quantph\undefined
\else
% Automatische Definition der Variablen
\qpvar{\qpno}
\qpvar{\qptitle}
\qpvar{\qpdate}
\qpvar{\qpcontribute}
\qpvar{\qpfirstdate}
\qpvar{\qplastdate}
\qpvar{\qpsolve}
\fi

%
% ***************************************************************************
%
% Variablenbelegungen in der LaTeX-Datei,
% einzufuegen durch IMaPh
%
% Diese Variablen sollten NICHT im TEXT-Teil
% definiert oder verwendet werden!
%

% Der folgende Block wird von TTH "verbatim" ausgegeben,
% alle anderen verarbeiten ihn
%%tth: \begin{html}
\qpno{12}
\qptitle{Bell Inequalities for long range vacuum correlations}
\qpdate{04 Apr 2005}
\qpcontribute{\href{http://www.imaph.tu-bs.de/qi/links.html}{R. Verch}}
\qpfirstdate{22 Jan 2002}
\qplastdate{\nothing}
\qpsolve{\nothing}
%%tth: \end{html}

%
% Für Vorschau-Version nur das nötigste einstellen
%
% \nonstopmode kann ggf. auskommentiert werden!
%
\else
\nonstopmode
\documentclass[12pt,a4paper]{article}
\usepackage[latin1]{inputenc}
\markboth{\centerline{\textsl{\bfseries Draft of QI open problem -- content test only!}}}{\centerline{\textsl{\bfseries Draft of QI open problem -- content test only!}}}
\pagestyle{myheadings}
\title{Draft of QI open problem -- content test only!}
\author{No page formatting}
\fi

%
% ***************************************************************************
%
% Praeambel, redigiert durch IMaPh
%

% Keine AMS-Erweiterungen!
% Standardbefehle (\text{}, \R, \C) sind bereits in qiheader.tex
% definiert.

% TTH fuegt Label "BEGINDOCUMENT" ein und ignoriert den Block
%%tth: \iftth BEGINDOCUMENT \begin{document} \else
\ifx\quantph\undefined
	\begin{document}
\fi
\maketitle
% \makeinfo
%%tth: \fi

%
% ***************************************************************************
%
% Text-Teil, redigiert durch IMaPh
%

%
% NICHT VERGESSEN:
%	- ggf. $...$ durch \(...\) ersetzen
%	- Eintraege ins Literaturverzeichnis vollstaendig klammern: \bibitem[ABC]{12:ABC}{A.A.,B.B., ...}
%

\section{Problem}

It is well known \cite{12:SW} that vacuum fluctuations maximally
violate  the CHSH-Bell inequalities for suitable spacelike separated
observables, and that this violation goes to zero as the two localization
regions are moved apart.

Decide whether some (necessarily small) violation of the
inequalities is possible for regions arbitrarily far apart. For
definiteness, consider a massive scalar free relativistic Bose
field.

\section{Background}

It is known \cite{12:HC} that the vacuum is not separable at any
distance.  More recently \cite{12:VW}, it has been shown that an
analogue of the  ``positive partial transpose''  condition fails
for  arbitrary regions at any distance. But the problem as stated above
remains open.

\ifx\quantph\undefined
	\end{document}
\fi

%\newpage\centerline{}

%
% Schalter für End-Version: \let\final\relax
% Schalter für Vorschau-Version: \final nicht definiert
%
\ifx\final\relax

%
% Header-Datei laden (voller Pfad, kein Endung ".tex")
%

\ifx\quantph\undefined
\else
% Automatische Definition der Variablen
\qpvar{\qpno}
\qpvar{\qptitle}
\qpvar{\qpdate}
\qpvar{\qpcontribute}
\qpvar{\qpfirstdate}
\qpvar{\qplastdate}
\qpvar{\qpsolve}
\fi

%
% ***************************************************************************
%
% Variablenbelegungen in der LaTeX-Datei,
% einzufuegen durch IMaPh
%
% Diese Variablen sollten NICHT im TEXT-Teil
% definiert oder verwendet werden!
%

% Der folgende Block wird von TTH "verbatim" ausgegeben,
% alle anderen verarbeiten ihn
%%tth: \begin{html}
\qpno{13}
\qptitle{Mutually unbiased bases}
\qpdate{04 Apr 2005}
\qpcontribute{\href{http://www.imaph.tu-bs.de/qi/links.html}{B.-G. Englert}}
\qpfirstdate{31 Jan 2003}
\qplastdate{07 Jan 2004}
\qpsolve{\nothing}
%%tth: \end{html}

%
% Für Vorschau-Version nur das nötigste einstellen
%
% \nonstopmode kann ggf. auskommentiert werden!
%
\else
\nonstopmode
\documentclass[12pt,a4paper]{article}
\usepackage[latin1]{inputenc}
\markboth{\centerline{\textsl{\bfseries Draft of QI open problem -- content test only!}}}{\centerline{\textsl{\bfseries Draft of QI open problem -- content test only!}}}
\pagestyle{myheadings}
\title{Draft of QI open problem -- content test only!}
\author{No page formatting}
\fi

%
% ***************************************************************************
%
% Praeambel, redigiert durch IMaPh
%

% Keine AMS-Erweiterungen!
% Standardbefehle (\text{}, \R, \C) sind bereits in qiheader.tex
% definiert.

%-ltoh-	:b/e	:\begin{quotation}:<blockquote>:</blockquote>:

% TTH fuegt Label "BEGINDOCUMENT" ein und ignoriert den Block
%%tth: \iftth BEGINDOCUMENT \begin{document} \else
\ifx\quantph\undefined
	\begin{document}
\fi
\maketitle
% \makeinfo
%%tth: \fi

%
% ***************************************************************************
%
% Text-Teil, redigiert durch IMaPh
%

%
% NICHT VERGESSEN:
%	- ggf. $...$ durch \(...\) ersetzen
%	- Eintraege ins Literaturverzeichnis vollstaendig klammern: \bibitem[ABC]{13:ABC}{A.A.,B.B., ...}
%

\section{Problem}

Determine the maximal number $K$ of orthonormal bases in a $D$-dimensional
Hilbert space, which are mutually unbiased in the following sense: If  $e_i^k$
denotes the $i$th vector of the $k$th basis, all scalar products $\langle
e_i^k,e_j^n\rangle$ with $k\neq n$ have the same absolute value (namely
$D^{-1/2}$).

It is known that if $D$ is the power of a prime, $K=D+1$ can be reached, but
this is not known for any other composite number. So the problem is already to
decide whether there exist $K=7$ mutually unbiased bases in $D=6$ dimensions.

\section{Background}

The problem comes up in at least three (related) contexts:

\begin{enumerate}
\renewcommand{\labelenumi}{(\arabic{enumi})}

\item \textbf{State determination} \cite{13:Iv}, \cite{13:WF}

Suppose we want to determine the density operator of a source by measuring $K$
observables (with $D$ one-dimensional projections each). Each such measurement
allows us to determine $D-1$ independent parameters, so we can determine
$K(D-1)$ out of $D^2-1$ parameters in the density operator. Hence $K=D+1$
should suffice. In order to achieve best estimation results, the measurements
should duplicate no information already contained in other measurements,
i.\,e., the observables should be pairwise complementary, or the bases mutually
unbiased in the above sense.

\item \textbf{Cryptography}

Suppose Alice sends $D$-level systems prepared in one of the $D$ pure states
$e_i^k$ belonging to a set of $K$ orthonormal bases agreed between Alice and
Bob. If Bob measures in the same basis, he can decode the value $i$ perfectly.
In cryptography one also wants that if an eavesdropper measures the system in
any one of the other bases, she can extract no information whatsoever about
$i$. This requires the bases to be mutually unbiased.

It is known that a higher error level can be tolerated in the channel for
protocols using maximal families of mutually unbiased bases (e.\,g., the ``six
state protocol'', $D=2$, $K=3$) rather than non-maximal ones (e.\,g., BB84,
using $D=2$, $K=2$).

\item \textbf{The Mean King} \cite{13:AE}

\begin{quotation} A ship-wrecked physicist gets stranded on a far-away island that
is ruled by a mean king who loves cats and hates physicists since the day when
he first heard what happened to Schrödinger's cat. A similar fate is awaiting
the stranded physicist. Yet, mean as he is, the king enjoys defeating
physicists on their own turf, and therefore he maliciously offers an apparently
virtual chance of rescue.

He takes the physicist to the royal laboratory, a splendid place where
experiments of any kind can be performed perfectly. There the king invites the
physicist to prepare a certain silver atom in any state she likes. The king's
men will then measure one of the three cartesian spin components of this atom
-- they'll either measure $\sigma_x$, $\sigma_y$, or $\sigma_z$ without,
however, telling the physicist which one of the measurements is actually done.
Then it is again the physicist's turn, and she can perform any experiment of
her choosing. Only after she's finished with it, the king will tell her which
spin component had been measured by his men. To save her neck, the physicist
must then state correctly the measurement result that the king's men had
obtained.

Much to the king's frustration, the physicist rises to the challenge -- and not
just by sheer luck: She gets the right answer any time the whole procedure is
repeated. How does she do it? \end{quotation}

More generally, the king's men might be allowed to perform one out of $K$ complete
von Neumann measurements on a $D$-dimensional system. The problem first came up in
\cite{13:VA+}, together with a solution for $D=2$. Solutions involving mutually
unbiased bases are presented in \cite{13:AE}, \cite{13:Ara}, \cite{13:Arb},
\cite{13:EA}.
Confer also the experimental realization in \cite{13:SS+}.
\end{enumerate}

\section{Partial Results}

% \cite{13:WF}
%
H. Barnum \cite{13:Ba} points out a close connection of this problem with ``spherical
2-designs'', which are collections of pure states such that the average of a
polynomial of degree 2 on these states equals the integral of the polynomial over
all pure states.

For the case $D=6$ there are a number of different but equivalent formulations of
this problem.

A. Pittenger and M. Rubin \cite{13:PR} give a constructive proof for the case of prime
power dimension \cite{13:WF}. They also adress the question of separabilty and provide
an appendix on the necessary parts of algebraic field extensions. Another proof can
be found in \cite{13:KR}.

C. Archer \cite{13:Arc} shows that even generalizations of these constructions do not
extend the results beyond prime power dimension.

\ifx\quantph\undefined
	\end{document}
\fi

%\newpage\centerline{}

%
% Schalter für End-Version: \let\final\relax
% Schalter für Vorschau-Version: \final nicht definiert
%
\ifx\final\relax

%
% Header-Datei laden (voller Pfad, kein Endung ".tex")
%

\ifx\quantph\undefined
\else
% Automatische Definition der Variablen
\qpvar{\qpno}
\qpvar{\qptitle}
\qpvar{\qpdate}
\qpvar{\qpcontribute}
\qpvar{\qpfirstdate}
\qpvar{\qplastdate}
\qpvar{\qpsolve}
\fi

%
% ***************************************************************************
%
% Variablenbelegungen in der LaTeX-Datei,
% einzufuegen durch IMaPh
%
% Diese Variablen sollten NICHT im TEXT-Teil
% definiert oder verwendet werden!
%

% Der folgende Block wird von TTH "verbatim" ausgegeben,
% alle anderen verarbeiten ihn
%%tth: \begin{html}
\qpno{14}
\qptitle{Tough error models}
\qpdate{04 Apr 2005}
\qpcontribute{\href{http://www.imaph.tu-bs.de/qi/links.html}{E. Knill}}
\qpfirstdate{31 Jan 2003}
\qplastdate{\nothing}
\qpsolve{\nothing}
%%tth: \end{html}

%
% Für Vorschau-Version nur das nötigste einstellen
%
% \nonstopmode kann ggf. auskommentiert werden!
%
\else
\nonstopmode
\documentclass[12pt,a4paper]{article}
\usepackage[latin1]{inputenc}
\markboth{\centerline{\textsl{\bfseries Draft of QI open problem -- content test only!}}}{\centerline{\textsl{\bfseries Draft of QI open problem -- content test only!}}}
\pagestyle{myheadings}
\title{Draft of QI open problem -- content test only!}
\author{No page formatting}
\fi

%
% ***************************************************************************
%
% Praeambel, redigiert durch IMaPh
%

% Keine AMS-Erweiterungen!
% Standardbefehle (\text{}, \R, \C, ...) sind bereits in qiheader.tex
% definiert.

% TTH fuegt Label "BEGINDOCUMENT" ein und ignoriert den Block
%%tth: \iftth BEGINDOCUMENT \begin{document} \else
\ifx\quantph\undefined
	\begin{document}
\fi
\maketitle
% \makeinfo
%%tth: \fi

%
% ***************************************************************************
%
% Text-Teil, redigiert durch IMaPh
%

%
% NICHT VERGESSEN:
%	- ggf. $...$ durch \(...\) ersetzen
%	- Eintraege ins Literaturverzeichnis vollstaendig klammern: \bibitem[ABC]{14:ABC}{A.A.,B.B., ...}
%

\section{Problem}
An error model $E$ is an $e$-dimensional vector space
of operators acting on an $n$-dimensional Hilbert space $H$. A
quantum code is a subspace $C\subset H$, and is said to correct
$E$, if the projector $P_C$ onto $C$ satisfies $P_C A^{*} B P_C =
\lambda(A,B) P_C$ for all $A,B\in E$, and suitable scalars
$\lambda(A,B)$.

\begin{itemize}
\item Given $e$ and $n$, find the largest $c=c(e,n)$ such that we
can assert the existence of a code $C$ of dimension $c$ without
further information about $E$.
\item Find ``tough error models'' for which this bound is (nearly)
tight.
\end{itemize}

\section{Background}
For an introduction to quantum error-correction see, for example, \cite{14:KL02}.

\section{Partial results}
See \cite{14:KL00}, where a lower bound of $c(e,n) > n/(e^{2} (e^{2}+1))$ is
given.

A trivial upper bound on $c(e,n)$ comes from taking orthogonal
projections of roughly equal dimension $n/e$ as the error model.
Since the channel with these Kraus operators (a L\"{u}ders-von
Neumann projective measurement) has capacity at most $n/e$, it is
impossible to find larger code spaces. Hence $c(e,n)\leq \lceil
n/e\rceil$.

\ifx\quantph\undefined
	\end{document}
\fi

%
% Schalter für End-Version: \let\final\relax
% Schalter für Vorschau-Version: \final nicht definiert
%
\ifx\final\relax

%
% Header-Datei laden (voller Pfad, kein Endung ".tex")
%

\ifx\quantph\undefined
\else
% Automatische Definition der Variablen
\qpvar{\qpno}
\qpvar{\qptitle}
\qpvar{\qpdate}
\qpvar{\qpcontribute}
\qpvar{\qpfirstdate}
\qpvar{\qplastdate}
\qpvar{\qpsolve}
\fi

%
% ***************************************************************************
%
% Variablenbelegungen in der LaTeX-Datei,
% einzufuegen durch IMaPh
%
% Diese Variablen sollten NICHT im TEXT-Teil
% definiert oder verwendet werden!
%

% Der folgende Block wird von TTH "verbatim" ausgegeben,
% alle anderen verarbeiten ihn
%%tth: \begin{html}
\qpno{15}
\qptitle{Separability from spectrum}
\qpdate{04 Apr 2005}
\qpcontribute{\href{http://www.imaph.tu-bs.de/qi/links.html}{E. Knill}}
\qpfirstdate{31 Jan 2003}
\qplastdate{13 Aug 2003}
\qpsolve{\nothing}
%%tth: \end{html}

%
% Für Vorschau-Version nur das nötigste einstellen
%
% \nonstopmode kann ggf. auskommentiert werden!
%
\else
\nonstopmode
\documentclass[12pt,a4paper]{article}
\usepackage[latin1]{inputenc}
\markboth{\centerline{\textsl{\bfseries Draft of QI open problem -- content test only!}}}{\centerline{\textsl{\bfseries Draft of QI open problem -- content test only!}}}
\pagestyle{myheadings}
\title{Draft of QI open problem -- content test only!}
\author{No page formatting}
\fi

%
% ***************************************************************************
%
% Praeambel, redigiert durch IMaPh
%

% Keine AMS-Erweiterungen!
% Standardbefehle (\text{}, \R, \C, ...) sind bereits in qiheader.tex
% definiert.

% TTH fuegt Label "BEGINDOCUMENT" ein und ignoriert den Block
%%tth: \iftth BEGINDOCUMENT \begin{document} \else
\ifx\quantph\undefined
	\begin{document}
\fi
\maketitle
% \makeinfo
%%tth: \fi

%
% ***************************************************************************
%
% Text-Teil, redigiert durch IMaPh
%

%
% NICHT VERGESSEN:
%	- ggf. $...$ durch \(...\) ersetzen
%	- Eintraege ins Literaturverzeichnis vollstaendig klammern: \bibitem[ABC]{15:ABC}{A.A.,B.B., ...}
%

\section{Problem}
For a mixed state $\rho$ on an $N M$-dimensional Hilbert space: Are there any
factorizations into an $N$ tensor an $M$ dimensional space with
respect to which the state is not seperable?  This depends only on the
spectrum of $\rho$ and the problem is to characterize the spectra for
which the answer is "no".

\section{Background}
The question arises in the context where
we are given a highly mixed state on two quantum systems and the ability
to apply any unitary operator. Can an inseperable state be obtained?
For sufficiently mixed states, this is not possible.

This problem is different from \hyperlink{Outline9.0}{No.~9}, because only the
spectrum of $\rho$ and not the spectra of the reductions are to be part of the
criterion.

\section{Partial results}
See the generic bounds on how close a state has to be
to the completely mixed state to be guaranteed not to have
entanglement. The paper of Leonid Gurvits and Howard Barnum \cite{15:GB02} has
further relevant results.

For the case of two qubits, the question is solved in \cite{15:VA01}: Exactly the
states with eigenvalues $x_1, x_2, x_3, x_4$ (arranged in decreasing order) obeying
$x_1 - x_3 - 2 \sqrt{x_2 x_4} \le 0$ cannot be transformed into a state with
non-zero entanglement of formation by applying any unitary operator (Theorem 1).

\section{Source}
Howard Barnum, Leonid Gurvits, E. K.

\ifx\quantph\undefined
	\end{document}
\fi

%
% Schalter für End-Version: \let\final\relax
% Schalter für Vorschau-Version: \final nicht definiert
%
\ifx\final\relax

%
% Header-Datei laden (voller Pfad, kein Endung ".tex")
%

\ifx\quantph\undefined
\else
% Automatische Definition der Variablen
\qpvar{\qpno}
\qpvar{\qptitle}
\qpvar{\qpdate}
\qpvar{\qpcontribute}
\qpvar{\qpfirstdate}
\qpvar{\qplastdate}
\qpvar{\qpsolve}
\fi

%
% ***************************************************************************
%
% Variablenbelegungen in der LaTeX-Datei,
% einzufuegen durch IMaPh
%
% Diese Variablen sollten NICHT im TEXT-Teil
% definiert oder verwendet werden!
%

% Der folgende Block wird von TTH "verbatim" ausgegeben,
% alle anderen verarbeiten ihn
%%tth: \begin{html}
\qpno{16}
\qptitle{Complexity of product preparations}
\qpdate{04 Apr 2005}
\qpcontribute{\href{http://www.imaph.tu-bs.de/qi/links.html}{E. Knill}}
\qpfirstdate{31 Jan 2003}
\qplastdate{\nothing}
\qpsolve{\nothing}
%%tth: \end{html}

%
% Für Vorschau-Version nur das nötigste einstellen
%
% \nonstopmode kann ggf. auskommentiert werden!
%
\else
\nonstopmode
\documentclass[12pt,a4paper]{article}
\usepackage[latin1]{inputenc}
\markboth{\centerline{\textsl{\bfseries Draft of QI open problem -- content test only!}}}{\centerline{\textsl{\bfseries Draft of QI open problem -- content test only!}}}
\pagestyle{myheadings}
\title{Draft of QI open problem -- content test only!}
\author{No page formatting}
\fi

%
% ***************************************************************************
%
% Praeambel, redigiert durch IMaPh
%

% Keine AMS-Erweiterungen!
% Standardbefehle (\text{}, \R, \C, ...) sind bereits in qiheader.tex
% definiert.
\providecommand{\ket}[1]{|#1\rangle}

% TTH fuegt Label "BEGINDOCUMENT" ein und ignoriert den Block
%%tth: \iftth BEGINDOCUMENT \begin{document} \else
\ifx\quantph\undefined
	\begin{document}
\fi
\maketitle
% \makeinfo
%%tth: \fi

%
% ***************************************************************************
%
% Text-Teil, redigiert durch IMaPh
%

%
% NICHT VERGESSEN:
%	- ggf. $...$ durch \(...\) ersetzen
%	- Eintraege ins Literaturverzeichnis vollstaendig klammern: \bibitem[ABC]{16:ABC}{A.A.,B.B., ...}
%

\section{Problem}
What can be said about the algorithmic complexity of preparing
$\ket{\psi}\ket{\psi}\ldots$ ($n$ times), asymptotically, as a function
of $n$ and the algorithmic complexity of preparing $\ket{\psi}$?  Take
$\ket{\psi}$ to be a state of $m$ qubits. By algorithmc complexity I
mean the number of gates required to prepare the state from
$\ket{0}$. This depends on the gate set used so the question concerns
asymptotics. For the present purposes, one can take as a
gate set all roations $e^{i\phi\sigma_u}$ where $\sigma_u$ is a
product of Pauli matrices. The complexity of this gate is $|\phi|$.
It might be useful to consider a version of this question involving an
approximation parameter also.

\section{Remark}
It may be possible to clone $\ket{\psi}$ more efficiently
than to prepare it, given that one knows $\ket{\psi}$. 

\section{Source}
E. K., Gerardo Ortiz, Rolando Somma.

\ifx\quantph\undefined
	\end{document}
\fi

%\newpage\centerline{}

%
% Schalter für End-Version: \let\final\relax
% Schalter für Vorschau-Version: \final nicht definiert
%
\ifx\final\relax

%
% Header-Datei laden (voller Pfad, kein Endung ".tex")
%

\ifx\quantph\undefined
\else
% Automatische Definition der Variablen
\qpvar{\qpno}
\qpvar{\qptitle}
\qpvar{\qpdate}
\qpvar{\qpcontribute}
\qpvar{\qpfirstdate}
\qpvar{\qplastdate}
\qpvar{\qpsolve}
\fi

%
% ***************************************************************************
%
% Variablenbelegungen in der LaTeX-Datei,
% einzufuegen durch IMaPh
%
% Diese Variablen sollten NICHT im TEXT-Teil
% definiert oder verwendet werden!
%

% Der folgende Block wird von TTH "verbatim" ausgegeben,
% alle anderen verarbeiten ihn
%%tth: \begin{html}
\qpno{17}
\qptitle{Reversibility of entanglement assisted coding}
\qpdate{04 Apr 2005}
\qpcontribute{\href{http://www.imaph.tu-bs.de/qi/links.html}{P. Shor}}
\qpfirstdate{31 Jan 2003}
\qplastdate{\nothing}
\qpsolve{\nothing}
%%tth: \end{html}

%
% Für Vorschau-Version nur das nötigste einstellen
%
% \nonstopmode kann ggf. auskommentiert werden!
%
\else
\nonstopmode
\documentclass[12pt,a4paper]{article}
\usepackage[latin1]{inputenc}
\markboth{\centerline{\textsl{\bfseries Draft of QI open problem -- content test only!}}}{\centerline{\textsl{\bfseries Draft of QI open problem -- content test only!}}}
\pagestyle{myheadings}
\title{Draft of QI open problem -- content test only!}
\author{No page formatting}
\fi

%
% ***************************************************************************
%
% Praeambel, redigiert durch IMaPh
%

% Keine AMS-Erweiterungen!
% Standardbefehle (\text{}, \R, \C) sind bereits in qiheader.tex
% definiert.

% TTH fuegt Label "BEGINDOCUMENT" ein und ignoriert den Block
%%tth: \iftth BEGINDOCUMENT \begin{document} \else
\ifx\quantph\undefined
	\begin{document}
\fi
\maketitle
% \makeinfo
%%tth: \fi

%
% ***************************************************************************
%
% Text-Teil, redigiert durch IMaPh
%

%
% NICHT VERGESSEN:
%	- ggf. $...$ durch \(...\) ersetzen
%	- Eintraege ins Literaturverzeichnis vollstaendig klammern: \bibitem[ABC]{17:ABC}{A.A.,B.B., ...}
%

\section{Problem}

For any two quantum channels $S$ and $T$, define the \emph{entanglement
assisted capacity} $C_\text{E}(T,S)$ of $T$ for $S$-messages as the supremum of
all rates $r$ such that, for large $n$, $rn$ parallel copies of $T$ may be
simulated by $n$ copies of $S$, where the simulation involves arbitrary coding
and decoding operations using (if necessary) arbitrarily many entangled pairs
between sender and receiver, and where the errors go to zero as $n\to\infty$.

Show that $C_\text{E}(T,S)=C_\text{E}(S,T)^{-1}$.

\section{Background}

As for other capacities, the \emph{two-step coding inequality} $C_\text{E}(T,S)
\,C_\text{E}(S,R)\leq C_\text{E}(T,R)$ is easy to show. Hence $C_\text{E}(T,S)
\,C_\text{E}(S,T)\leq1$. Equality means here, that the two channels are
essentially equivalent as a resource for simulating other channels $R$ (apart
from a constant factor): $C_\text{E}(R,S)=\text{const}\, C_\text{E}(R,T)$ (with
$\text{const}=C_\text{E}(T,S)$). In this case we call $S$ and $T$ reversible
for entanglement assisted coding.

For ordinary capacity $C(T,S)$ (without entanglement assistance) reversibility
fails in general: When $S$ is an ideal classical 1 bit channel, and $T$ is an
ideal 1 qubit quantum channel, we have $C(S,T)=1$, but $C(T,S)=0$, because
quantum information cannot be sent on classical channels. On the other hand,
with entanglement assistance we have $C(S,T)=2$ by superdense coding and
$C(T,S)=1/2$ by teleportation.

Because all ideal channels $S$ are equivalent as reference channels, we can
define $C_\text{E}(T)=C_\text{E}(T,S_1)$, with $S_1$ the ideal classical 1 bit
channel as the entanglement assisted capacity of $T$. For this quantity there
is an explicit formula (coding theorem) by \cite{17:BSST1}. The problem stated
above appears in \cite{17:BSST2} as the ``Reverse Shannon Theorem''.

\section{Partial Solutions}

The problem is solved for the special case of a known ``tensor power source'',
i.\,e. a source emitting the same, known, density matrix at each time step.
Recent efforts by P. Shor focus on the unknown tensor power source and the
known ``tensor product source'' where the density matrix of the source is a
tensor product \cite{17:SH}.

\ifx\quantph\undefined
	\end{document}
\fi

%
% Schalter für End-Version: \let\final\relax
% Schalter für Vorschau-Version: \final nicht definiert
%
\ifx\final\relax

%
% Header-Datei laden (voller Pfad, kein Endung ".tex")
%

\ifx\quantph\undefined
\else
% Automatische Definition der Variablen
\qpvar{\qpno}
\qpvar{\qptitle}
\qpvar{\qpdate}
\qpvar{\qpcontribute}
\qpvar{\qpfirstdate}
\qpvar{\qplastdate}
\qpvar{\qpsolve}
\fi

%
% ***************************************************************************
%
% Variablenbelegungen in der LaTeX-Datei,
% einzufuegen durch IMaPh
%
% Diese Variablen sollten NICHT im TEXT-Teil
% definiert oder verwendet werden!
%

% Der folgende Block wird von TTH "verbatim" ausgegeben,
% alle anderen verarbeiten ihn
%%tth: \begin{html}
\qpno{18}
\qptitle{Qubit bi-negativity}
\qpdate{04 Apr 2005}
\qpcontribute{\href{http://www.imaph.tu-bs.de/qi/links.html}{K. G. H. Vollbrecht}}
\qpfirstdate{10 Feb 2003}
\qplastdate{\nothing}
\qpsolve{\nothing}
%%tth: \end{html}

%
% Für Vorschau-Version nur das nötigste einstellen
%
% \nonstopmode kann ggf. auskommentiert werden!
%
\else
\nonstopmode
\documentclass[12pt,a4paper]{article}
\usepackage[latin1]{inputenc}
\markboth{\centerline{\textsl{\bfseries Draft of QI open problem -- content test only!}}}{\centerline{\textsl{\bfseries Draft of QI open problem -- content test only!}}}
\pagestyle{myheadings}
\title{Draft of QI open problem -- content test only!}
\author{No page formatting}
\fi

%
% ***************************************************************************
%
% Praeambel, redigiert durch IMaPh
%

% Keine AMS-Erweiterungen!
% Standardbefehle (\text{}, \R, \C) sind bereits in qiheader.tex
% definiert.

% TTH fuegt Label "BEGINDOCUMENT" ein und ignoriert den Block
%%tth: \iftth BEGINDOCUMENT \begin{document} \else
\ifx\quantph\undefined
	\begin{document}
\fi
\maketitle
% \makeinfo
%%tth: \fi

%
% ***************************************************************************
%
% Text-Teil, redigiert durch IMaPh
%

%
% NICHT VERGESSEN:
%	- ggf. $...$ durch \(...\) ersetzen
%	- Eintraege ins Literaturverzeichnis vollstaendig klammern: \bibitem[ABC]{18:ABC}{A.A.,B.B., ...}
%
\section{Problem}

A little problem introduced in \cite{18:AMVW02} is the bi-negativity on two
qubits: Prove that

\[ |\sigma^{T_2}|^{T_2}\ge 0 \]

holds for every two-qubit state $\sigma$. Here, $T_2$ denotes the partial
transpose with respect to the second system (see also
\hyperlink{Outline2.0}{problem 2}) and $|.|$ is the operator absolute value,
$|x| = \sqrt{x^*\,x}$.

\ifx\quantph\undefined
	\end{document}
\fi

%\newpage\centerline{}

%
% Schalter für End-Version: \let\final\relax
% Schalter für Vorschau-Version: \final nicht definiert
%
\ifx\final\relax

%
% Header-Datei laden (voller Pfad, kein Endung ".tex")
%

\ifx\quantph\undefined
\else
% Automatische Definition der Variablen
\qpvar{\qpno}
\qpvar{\qptitle}
\qpvar{\qpdate}
\qpvar{\qpcontribute}
\qpvar{\qpfirstdate}
\qpvar{\qplastdate}
\qpvar{\qpsolve}
\fi

%
% ***************************************************************************
%
% Variablenbelegungen in der LaTeX-Datei,
% einzufuegen durch IMaPh
%
% Diese Variablen sollten NICHT im TEXT-Teil
% definiert oder verwendet werden!
%

% Der folgende Block wird von TTH "verbatim" ausgegeben,
% alle anderen verarbeiten ihn
%%tth: \begin{html}
\qpno{19}
\qptitle{Stronger Bell Inequalities for Werner states?}
\qpdate{04 Apr 2005}
\qpcontribute{\href{http://www.imaph.tu-bs.de/qi/links.html}{N. Gisin}}
\qpfirstdate{20 Jun 2003}
\qplastdate{\nothing}
\qpsolve{\nothing}
%%tth: \end{html}

%
% Für Vorschau-Version nur das nötigste einstellen
%
% \nonstopmode kann ggf. auskommentiert werden!
%
\else
\nonstopmode
\documentclass[12pt,a4paper]{article}
\usepackage[latin1]{inputenc}
\markboth{\centerline{\textsl{\bfseries Draft of QI open problem -- content test only!}}}{\centerline{\textsl{\bfseries Draft of QI open problem -- content test only!}}}
\pagestyle{myheadings}
\title{Draft of QI open problem -- content test only!}
\author{No page formatting}
\fi

%
% ***************************************************************************
%
% Praeambel, redigiert durch IMaPh
%

% Keine AMS-Erweiterungen!
% Standardbefehle (\text{}, \R, \C) sind bereits in qiheader.tex
% definiert.

% TTH fuegt Label "BEGINDOCUMENT" ein und ignoriert den Block
%%tth: \iftth BEGINDOCUMENT \begin{document} \else
\ifx\quantph\undefined
	\begin{document}
\fi
\maketitle
% \makeinfo
%%tth: \fi

%
% ***************************************************************************
%
% Text-Teil, redigiert durch IMaPh
%

%
% NICHT VERGESSEN:
%	- ggf. $...$ durch \(...\) ersetzen
%	- Eintraege ins Literaturverzeichnis vollstaendig klammern: \bibitem[ABC]{19:ABC}{A.A.,B.B., ...}
%

\section{Problem}

Find Bell Inequalities which are stronger than the CHSH inequalities in the sense
that they are violated by a wider range of Werner states.

\section{Background}

Recently, Daniel Collins and Nicolas Gisin \cite{19:CG} found a Bell Inequality and
states which violate the new but not the CHSH inequalities. Alas, the range of
Werner states violating the new inequality is smaller than that for the CHSH
setting.

\ifx\quantph\undefined
	\end{document}
\fi

%\newpage\centerline{}

%
% Schalter für End-Version: \let\final\relax
% Schalter für Vorschau-Version: \final nicht definiert
%
\ifx\final\relax

%
% Header-Datei laden (voller Pfad, kein Endung ".tex")
%

\ifx\quantph\undefined
\else
% Automatische Definition der Variablen
\qpvar{\qpno}
\qpvar{\qptitle}
\qpvar{\qpdate}
\qpvar{\qpcontribute}
\qpvar{\qpfirstdate}
\qpvar{\qplastdate}
\qpvar{\qpsolve}
\fi

%
% ***************************************************************************
%
% Variablenbelegungen in der LaTeX-Datei,
% einzufuegen durch IMaPh
%
% Diese Variablen sollten NICHT im TEXT-Teil
% definiert oder verwendet werden!
%

% Der folgende Block wird von TTH "verbatim" ausgegeben,
% alle anderen verarbeiten ihn
%%tth: \begin{html}
\qpno{20}
\qptitle{Reversible entanglement manipulation}
\qpdate{04 Apr 2005}
\qpcontribute{\href{http://www.imaph.tu-bs.de/qi/links.html}{M. B. Plenio}}
\qpfirstdate{08 Feb 2005}
\qplastdate{\nothing}
\qpsolve{\nothing}
%%tth: \end{html}

%
% Für Vorschau-Version nur das nötigste einstellen
%
% \nonstopmode kann ggf. auskommentiert werden!
%
\else
\nonstopmode
\documentclass[12pt,a4paper]{article}
\usepackage[latin1]{inputenc}
\markboth{\centerline{\textsl{\bfseries Draft of QI open problem -- content test only!}}}{\centerline{\textsl{\bfseries Draft of QI open problem -- content test only!}}}
\pagestyle{myheadings}
\title{Draft of QI open problem -- content test only!}
\author{No page formatting}
\fi

%
% ***************************************************************************
%
% Praeambel, redigiert durch IMaPh
%

% Keine AMS-Erweiterungen!
% Standardbefehle (\text{}, \R, \C) sind bereits in qiheader.tex
% definiert.

% TTH fuegt Label "BEGINDOCUMENT" ein und ignoriert den Block
%%tth: \iftth BEGINDOCUMENT \begin{document} \else
\ifx\quantph\undefined
	\begin{document}
\fi
\maketitle
% \makeinfo
%%tth: \fi

%
% ***************************************************************************
%
% Text-Teil, redigiert durch IMaPh
%

%
% NICHT VERGESSEN:
%	- ggf. $...$ durch \(...\) ersetzen
%	- Eintraege ins Literaturverzeichnis vollstaendig klammern: \bibitem[ABC]{20:ABC}{A.A.,B.B., ...}
%

\section{Problem}
The concept of entanglement as a resource motivates the study of
its transformation properties under certain classes of operations
such as local operations and classical communication (LOCC).

For a finite number of identically prepared quantum systems the
manipulation of entanglement under LOCC is generally irreversible,
both for pure and mixed states. In the asymptotic limit of
infinitely many identical copies of a pure state, in contrast,
pure bi-partite entanglement can be interconverted reversibly
\cite{20:BBPS96}. For mixed states, however, this asymptotic
reversibility under LOCC operations is lost \cite{20:VC02,20:HSS02}.

However, there are more general sets of operations for which
entanglement manipulation might become reversible again. One such
example is the set of positive partial transpose preserving
operations (ppt-operations) \cite{20:Ra00} which are all those
completely positive maps that map the set of ppt-states into
itself. It has been shown that under ppt-operations there are some
mixed states that can be reversible converted into pure singlet
states in the asymptotic limit \cite{20:APE03}. This has
been proven for the totally anti-symmetric Werner state and weak
numerical evidence suggests that this is true for all Werner
states \cite{20:Pl}. On the other hand in \cite{20:HOH02}
it was shown that under certain conditions and for a set of
operations (denoted Hyper-set in \cite{20:HOH02}) that is
smaller than ppt-operations and strictly larger than LOCC
asymptotic irreversibility persists.

Asymptotic reversibility under a class of operations would lead to
a unique entanglement measure and impose a unique ordering on
entangled states thereby playing a role similar to entropy in
thermodynamics.

The following are open questions:
\begin{itemize}
	\item Are ppt-operations sufficient to ensure asymptotically
    	reversibly interconversion of \emph{all}, i.e. pure and mixed,
	    bi-partite entangled states \cite{20:BFC}?
    \item What is the smallest non-trivial class of operations that permits
        asymptotically reversibly interconversion of \emph{all}, i.e. pure and
        mixed, bi-partite entangled states \cite{20:Bet}?
\end{itemize}

\ifx\quantph\undefined
	\end{document}
\fi

%
% Schalter für End-Version: \let\final\relax
% Schalter für Vorschau-Version: \final nicht definiert
%
\ifx\final\relax

%
% Header-Datei laden (voller Pfad, kein Endung ".tex")
%

\ifx\quantph\undefined
\else
% Automatische Definition der Variablen
\qpvar{\qpno}
\qpvar{\qptitle}
\qpvar{\qpdate}
\qpvar{\qpcontribute}
\qpvar{\qpfirstdate}
\qpvar{\qplastdate}
\qpvar{\qpsolve}
\fi

%
% ***************************************************************************
%
% Variablenbelegungen in der LaTeX-Datei,
% einzufuegen durch IMaPh
%
% Diese Variablen sollten NICHT im TEXT-Teil
% definiert oder verwendet werden!
%

% Der folgende Block wird von TTH "verbatim" ausgegeben,
% alle anderen verarbeiten ihn
%%tth: \begin{html}
\qpno{21}
\qptitle{Bell violation by tensoring}
\qpdate{04 Apr 2005}
\qpcontribute{\href{http://www.qinfo.org/?page_id=23}{Y. C. Liang}}
\qpfirstdate{08 Feb 2005}
\qplastdate{\nothing}
\qpsolve{\nothing}
%%tth: \end{html}

%
% Für Vorschau-Version nur das nötigste einstellen
%
% \nonstopmode kann ggf. auskommentiert werden!
%
\else
\nonstopmode
\documentclass[12pt,a4paper]{article}
\usepackage[latin1]{inputenc}
\markboth{\centerline{\textsl{\bfseries Draft of QI open problem -- content test only!}}}{\centerline{\textsl{\bfseries Draft of QI open problem -- content test only!}}}
\pagestyle{myheadings}
\title{Draft of QI open problem -- content test only!}
\author{No page formatting}
\fi

%
% ***************************************************************************
%
% Praeambel, redigiert durch IMaPh
%

% Keine AMS-Erweiterungen!
% Standardbefehle (\text{}, \R, \C) sind bereits in qiheader.tex
% definiert.

% TTH fuegt Label "BEGINDOCUMENT" ein und ignoriert den Block
%%tth: \iftth BEGINDOCUMENT \begin{document} \else
\ifx\quantph\undefined
	\begin{document}
\fi
\maketitle
% \makeinfo
%%tth: \fi

%
% ***************************************************************************
%
% Text-Teil, redigiert durch IMaPh
%

%
% NICHT VERGESSEN:
%	- ggf. $...$ durch \(...\) ersetzen
%	- Eintraege ins Literaturverzeichnis vollstaendig klammern: \bibitem[ABC]{21:ABC}{A.A.,B.B., ...}
%

\section{Problem}

Can one find bipartite density operators $\rho_{1,2}$,
neither of which violates any CHSH Bell inequality, with the
property that $\rho_1\otimes\rho_2$ does?

\ifx\quantph\undefined
	\end{document}
\fi

%\newpage\centerline{}

%
% Schalter für End-Version: \let\final\relax
% Schalter für Vorschau-Version: \final nicht definiert
%
\ifx\final\relax

%
% Header-Datei laden (voller Pfad, kein Endung ".tex")
%

\ifx\quantph\undefined
\else
% Automatische Definition der Variablen
\qpvar{\qpno}
\qpvar{\qptitle}
\qpvar{\qpdate}
\qpvar{\qpcontribute}
\qpvar{\qpfirstdate}
\qpvar{\qplastdate}
\qpvar{\qpsolve}
\fi

%
% ***************************************************************************
%
% Variablenbelegungen in der LaTeX-Datei,
% einzufuegen durch IMaPh
%
% Diese Variablen sollten NICHT im TEXT-Teil
% definiert oder verwendet werden!
%

% Der folgende Block wird von TTH "verbatim" ausgegeben,
% alle anderen verarbeiten ihn
%%tth: \begin{html}
\qpno{22}
\qptitle{Asymptotic cloning is state estimation?}
\qpdate{04 Apr 2005}
\qpcontribute{\href{http://www.imaph.tu-bs.de/staff/mk.html}{M. Keyl}}
\qpfirstdate{10 Feb 2005}
\qplastdate{\nothing}
\qpsolve{\nothing}
%%tth: \end{html}

%
% Für Vorschau-Version nur das nötigste einstellen
%
% \nonstopmode kann ggf. auskommentiert werden!
%
\else
\nonstopmode
\documentclass[12pt,a4paper]{article}
\usepackage[latin1]{inputenc}
\markboth{\centerline{\textsl{\bfseries Draft of QI open problem -- content test only!}}}{\centerline{\textsl{\bfseries Draft of QI open problem -- content test only!}}}
\pagestyle{myheadings}
\title{Draft of QI open problem -- content test only!}
\author{No page formatting}
\fi

%
% ***************************************************************************
%
% Praeambel, redigiert durch IMaPh
%

% Keine AMS-Erweiterungen!
% Standardbefehle (\text{}, \R, \C) sind bereits in qiheader.tex
% definiert.

% TTH fuegt Label "BEGINDOCUMENT" ein und ignoriert den Block
%%tth: \iftth BEGINDOCUMENT \begin{document} \else
\ifx\quantph\undefined
	\begin{document}
\fi
\maketitle
% \makeinfo
%%tth: \fi

%
% ***************************************************************************
%
% Text-Teil, redigiert durch IMaPh
%

%
% NICHT VERGESSEN:
%	- ggf. $...$ durch \(...\) ersetzen
%	- Eintraege ins Literaturverzeichnis vollstaendig klammern: \bibitem[ABC]{22:ABC}{A.A.,B.B., ...}
%

\section{Problem}

Fix an arbitrary probability measure on the pure states of a $d$-dimensional
quantum system. Let $F(N,M)$ be the optimal single copy fidelity for $N$-to-$M$
cloning transformations, averaged with respect to the given probability measure
and over all $M$ clones.

On the other hand, let $F(N,\infty)$ be the best mean fidelity achievable by
measuring on $N$ input copies of the state, and repreparing a state according to
the measured data. The problem is to decide whether one always gets
\[ \lim_{M\to\infty} F(N,M) = F(N,\infty). \]

It is clear that the limit exists, because $F(N,M)$ is non-increasing in $M$.
Moreover, the limit will be larger or equal than the right hand side,  because
estimation with repreaparation is a particular cloning method. A weaker, but still
interesting version of the problem is whether the above equation becomes true in
the limit $N\to\infty$.

\section{Background}

In the examples \cite{22:KW99,22:BCDM00}, where optimal cloner and estimator have been
computed, the formula is true. The limit formula is a piece of
folklore, partly based on the idea that if one has many clones, one
could make a statistical measurement on them and thereby obtain a
good estimation. This reasoning is faulty, however, because it neglects the
correlations, and possibly the entanglement between the clones.

\ifx\quantph\undefined
	\end{document}
\fi

%
% Schalter für End-Version: \let\final\relax
% Schalter für Vorschau-Version: \final nicht definiert
%
\ifx\final\relax

%
% Header-Datei laden (voller Pfad, kein Endung ".tex")
%

\ifx\quantph\undefined
\else
% Automatische Definition der Variablen
\qpvar{\qpno}
\qpvar{\qptitle}
\qpvar{\qpdate}
\qpvar{\qpcontribute}
\qpvar{\qpfirstdate}
\qpvar{\qplastdate}
\qpvar{\qpsolve}
\fi

%
% ***************************************************************************
%
% Variablenbelegungen in der LaTeX-Datei,
% einzufuegen durch IMaPh
%
% Diese Variablen sollten NICHT im TEXT-Teil
% definiert oder verwendet werden!
%

% Der folgende Block wird von TTH "verbatim" ausgegeben,
% alle anderen verarbeiten ihn
%%tth: \begin{html}
\qpno{23}
\qptitle{SIC POVMs and Zauner's Conjecture}
\qpdate{04 Apr 2005}
\qpcontribute{\href{http://www.quantum.physik.uni-potsdam.de/members/davidg.html}{D. Gross}}
\qpfirstdate{17 Feb 2005}
\qplastdate{\nothing}
\qpsolve{\nothing}
%%tth: \end{html}

%
% Für Vorschau-Version nur das nötigste einstellen
%
% \nonstopmode kann ggf. auskommentiert werden!
%
\else
\nonstopmode
\documentclass[12pt,a4paper]{article}
\usepackage[latin1]{inputenc}
\markboth{\centerline{\textsl{\bfseries Draft of QI open problem -- content test only!}}}{\centerline{\textsl{\bfseries Draft of QI open problem -- content test only!}}}
\pagestyle{myheadings}
\title{Draft of QI open problem -- content test only!}
\author{No page formatting}
\fi

%
% ***************************************************************************
%
% Praeambel, redigiert durch IMaPh
%

% Keine AMS-Erweiterungen!
% Standardbefehle (\text{}, \R, \C) sind bereits in qiheader.tex
% definiert.
\providecommand{\Ket}[1]{|#1\rangle}
\providecommand{\Bra}[1]{\langle#1|}
\providecommand{\Braket}[2]{\langle#1|#2\rangle}
\providecommand{\tr}{\mathrm{tr}}
%-ltoh-	:{}		:\Ket:|:>:
%-ltoh-	:{}		:\Bra:<:|:
%-ltoh-	:{2}	:\Braket:<#1|#2>::

%-ltoh- :comm	:\ast:*::
%-ltoh-	:{}		:\mathbf:<B>:</B>:
%-ltoh-	:b/e	:\begin{equation}:<p align=center>:</p>:

%-ltoh-	:{}		:\subsection[*]?:<br><p><font size=+1><b>:</b></font>:

% TTH fuegt Label "BEGINDOCUMENT" ein und ignoriert den Block
%%tth: \iftth BEGINDOCUMENT \begin{document} \else
\ifx\quantph\undefined
	\begin{document}
\fi
\maketitle
% \makeinfo
%%tth: \fi

%
% ***************************************************************************
%
% Text-Teil, redigiert durch IMaPh
%

%
% NICHT VERGESSEN:
%	- ggf. $...$ durch \(...\) ersetzen
%	- Eintraege ins Literaturverzeichnis vollstaendig klammern: \bibitem[ABC]{23:ABC}{A.A.,B.B., ...}
%

\section{Problem}

We will give three variants of the problem, each being stronger than its
predecessor. The terminology of problems 1 and 2 is taken mainly from
\cite{23:1}. For problem 3 see \cite{23:2} and \cite{23:3}.

\subsection*{Problem 1: SIC-POVMs}

A set of $d^2$ normed vectors $\{\Ket{\phi_i}\}_i$ in a Hilbert space of dimension
$d$ constitutes a set of \emph{equiangular lines} if their mutual inner products 
\[
	\left|\Braket{\phi_i}{\phi_j}\right|^2
\]
are independent of the choice of $i \neq j$. It can be shown \cite{23:1} that 
\begin{itemize}
	\item the associated projection operators sum to a multiple of 
		unity and thus induce a POVM (up to normalization) and that
	\item these operators are linearly independent and hence any quantum
		state can be reconstructed from the measurement statistics
		$p_i :=	\tr\left(\Ket{\phi_i} \Bra{\phi_i} \rho \right)$ of the POVM.
\end{itemize}
A POVM that arises in this way is called \emph{symmetric informationally
complete}, or a \emph{SIC-POVM} for short.

The most general form of the problem is: decide if SIC-POVMs exists in any
dimension $d$.

\subsection*{Problem 2: Covariant SIC-POVMs}

For a given basis $\left\{\Ket{q}\right\}_{q=0\ldots d-1}$ of the Hilbert space,
define the \emph{shift operator} $X$ and \emph{clock operator} $Z$ respectively by
the relations
\begin{eqnarray*}
	X \Ket{q} &:=& \Ket{q+1} \\
	Z \Ket{q} &:=& e^{i \frac {2 \pi} d q} \Ket{q},
\end{eqnarray*}
where arithmetic is modulo $d$. Further, define the \emph{Weyl operators}
\begin{equation}\label{23:weyl}
  w(p,q) = Z(p) X(q)
\end{equation}
for all $p, q \in {\mathbf{Z}}_d$. We will refer to the group generated
by (\ref{23:weyl}) as the \emph{Heisenberg group}. It is also known as the 
\emph{Weyl-Heisenberg group} or \emph{Generalized Pauli group}.

A vector $\Ket{\phi}$ is called a \emph{fiducial vector} with respect to the
Heisenberg group if the set 
\begin{equation}\label{23:cov}
	\left\{w(p,q) \, \Ket{\phi} \Bra{\phi} \, w(p,q)^\ast \right\}_{p,q=0\ldots d-1}
\end{equation}
induces a SIC-POVM. Such a SIC-POVM is said to be \emph{group covariant}. The
definition makes sense for any group of order at least $d^2$. However, we will
focus on the Heisenberg group in what follows.

The problem: decide if group covariant SIC-POVMs exist in any dimension $d$. 

\subsection*{Problem 3: Zauner's Conjecture}

The normalizer of the Heisenberg group within the unitaries $U(d)$ is
called the \emph{Clifford group}. There exists an element $z$ of the
Clifford group which is defined via its action on the Weyl operators
as
\[
	z \,w(p,q) z^\ast = w(q-p,-p).
\]
Zauner's conjecture, as formulated in \cite{23:3}, runs: in any dimension $d$,
a fiducial vector can be found among the eigenvectors  of $Z$.

\section{Background}

Besides their mathematical appeal, SIC-POVMs have obvious
applications to quantum state tomography. The symmetry condition assures
that the possible measurement outcomes are in some sense maximally
complementary. 

\section{Partial Results and History}

\begin{itemize}

\item
In the context of quantum information, the problem seems to have been
tackled first by Gerhard Zauner in his doctorial thesis \cite{23:2}
in 1999. To our knowledge, the results were neither published nor
translated into English, which caused some confusion in the English
literature, as to what Zauner had actually conjectured\footnote{Refer e.g.
to the first vs. the second version of \cite{23:3} on the arXiv
server.}. Zauner analyzed the spectrum of $z$. He listed
analytical expressions for fiducial vectors in dimension 2, 3, 4, 5 and
numerical expressions for $d=6, 7$. He noted that for dimension 8 an
analytic SIC-POVM is known, which is covariant under the action of the 
threefold tensor product of the two dimensional Heisenberg group.

\item
Wide interest in the problem arose with the 2003 paper by Renes et.
al. \cite{23:1}. Building on concepts from \emph{frame theory}, the authors
reduced the task of numerically finding fiducial vectors to a non-convex
global optimization problem. Using this method, they presented
numerical fiducial vectors for all dimensions up to 45 and counted the
number of distinct covariant SIC-POVMs up to dimension 7. The question
of whether those vectors were eigenstates of a Clifford operation was left
open (but see below). 
Further, four groups other than the Heisenberg
group were numerically found to induce SIC-POVMs in the sense of
(\ref{23:cov}).

The authors showed that a SIC-POVM corresponds to a \emph{spherical
2-design}\footnote{A finite set $X$ of unit vectors is a \emph{t-design} if the
average of any t-th order polynomial over $X$ is the same as the average of
that polynomial over the entire unit sphere.}. The same assertion was proven by
Klappenecker and Rötteler in \cite{23:4} and was apparently known to
Zauner (see Remark 3 in \cite{23:4}).

\item 
In \cite{23:5} Grassl used a computer algebra system capable of
symbolic calculations to prove Zauner's conjecture for $d=6$. He
remarked that elements of the Clifford group map fiducial vectors onto
fiducial vectors. Building on that observation, he could account for all 96
covariant SIC-POVMs that were reported to exist for $d=6$ in
\cite{23:1}.

\item
Appleby in \cite{23:3} gave a detailed description of the Clifford
group and extended it by allowing for anti-unitary operators. 
He verified that the numeric solutions of \cite{23:1} were compatible
with Zauner's conjecture and analyzed their stability groups inside the 
Clifford group\footnote{A similar analysis can be performed using 
discrete Wigner functions, as will be reported in \cite{23:6}.}.
Appleby goes on to present
analytical expressions for fiducial vectors in dimension 7 and 19 and
specifies an infinite sequence of dimensions for which he conjectures that
solutions can be found more easily.

\item
Inspired by a construction that links finite geometries to MUBs,
there have been some speculations by Wootters about whether SIC-POVMs
can be linked to finite affine planes \cite{23:7}. 
The same line of thought was pursued by Bengtsson and Ericsson in
\cite{23:8}. 
However, the existence of such a construction remains an open problem. 
The results by Grassl are of some relevance here, as it is known that
affine planes of order 6 do not exist.

\end{itemize}

\ifx\quantph\undefined
	\end{document}
\fi

%
% Schalter für End-Version: \let\final\relax
% Schalter für Vorschau-Version: \final nicht definiert
%
\ifx\final\relax

%
% Header-Datei laden (voller Pfad, kein Endung ".tex")
%

\ifx\quantph\undefined
\else
% Automatische Definition der Variablen
\qpvar{\qpno}
\qpvar{\qptitle}
\qpvar{\qpdate}
\qpvar{\qpcontribute}
\qpvar{\qpfirstdate}
\qpvar{\qplastdate}
\qpvar{\qpsolve}
\fi

%
% ***************************************************************************
%
% Variablenbelegungen in der LaTeX-Datei,
% einzufuegen durch IMaPh
%
% Diese Variablen sollten NICHT im TEXT-Teil
% definiert oder verwendet werden!
%

% Der folgende Block wird von TTH "verbatim" ausgegeben,
% alle anderen verarbeiten ihn
%%tth: \begin{html}
\qpno{24}
\qptitle{Secret key from all entangled states}
\qpdate{04 Apr 2005}
\qpcontribute{\href{http://www.imaph.tu-bs.de/qi/links.html}{P. Horodecki}}
\qpfirstdate{15 Mar 2005}
\qplastdate{\nothing}
\qpsolve{\nothing}
%%tth: \end{html}

%
% Für Vorschau-Version nur das nötigste einstellen
%
% \nonstopmode kann ggf. auskommentiert werden!
%
\else
\nonstopmode
\documentclass[12pt,a4paper]{article}
\usepackage[latin1]{inputenc}
\markboth{\centerline{\textsl{\bfseries Draft of QI open problem -- content test only!}}}{\centerline{\textsl{\bfseries Draft of QI open problem -- content test only!}}}
\pagestyle{myheadings}
\title{Draft of QI open problem -- content test only!}
\author{No page formatting}
\fi

%
% ***************************************************************************
%
% Praeambel, redigiert durch IMaPh
%

% Keine AMS-Erweiterungen!
% Standardbefehle (\text{}, \R, \C) sind bereits in qiheader.tex
% definiert.

% TTH fuegt Label "BEGINDOCUMENT" ein und ignoriert den Block
%%tth: \iftth BEGINDOCUMENT \begin{document} \else
\ifx\quantph\undefined
	\begin{document}
\fi
\maketitle
% \makeinfo
%%tth: \fi

%
% ***************************************************************************
%
% Text-Teil, redigiert durch IMaPh
%

%
% NICHT VERGESSEN:
%	- ggf. $...$ durch \(...\) ersetzen
%	- Eintraege ins Literaturverzeichnis vollstaendig klammern: \bibitem[ABC]{24:ABC}{A.A.,B.B., ...}
%

\section{Problem}

Can all bipartite entangled states be used to generate secrete
keys?

\section{Background}

In \cite{24:HHHO02} it is shown that some bound entangled states do
allow the extraction of a secret key. This is an extreme
counterexample to the idea that secret key is best generated from
an entangled state by first distilling pure singlets, and using
these to get the key. In principle, this provides a new
distinction among bipartite states in those which allow key
generation and those which do not. The problem asks whether this is
really a new distinction.

\ifx\quantph\undefined
	\end{document}
\fi

%\newpage\centerline{}

%
% Schalter für End-Version: \let\final\relax
% Schalter für Vorschau-Version: \final nicht definiert
%
\ifx\final\relax

%
% Header-Datei laden (voller Pfad, kein Endung ".tex")
%

\ifx\quantph\undefined
\else
% Automatische Definition der Variablen
\qpvar{\qpno}
\qpvar{\qptitle}
\qpvar{\qpdate}
\qpvar{\qpcontribute}
\qpvar{\qpfirstdate}
\qpvar{\qplastdate}
\qpvar{\qpsolve}
\fi

%
% ***************************************************************************
%
% Variablenbelegungen in der LaTeX-Datei,
% einzufuegen durch IMaPh
%
% Diese Variablen sollten NICHT im TEXT-Teil
% definiert oder verwendet werden!
%

% Der folgende Block wird von TTH "verbatim" ausgegeben,
% alle anderen verarbeiten ihn
%%tth: \begin{html}
\qpno{25}
\qptitle{Lockable entanglement measures}
\qpdate{04 Apr 2005}
\qpcontribute{\href{http://www.imaph.tu-bs.de/qi/links.html}{P. Horodecki}}
\qpfirstdate{15 Mar 2005}
\qplastdate{\nothing}
\qpsolve{\nothing}
%%tth: \end{html}

%
% Für Vorschau-Version nur das nötigste einstellen
%
% \nonstopmode kann ggf. auskommentiert werden!
%
\else
\nonstopmode
\documentclass[12pt,a4paper]{article}
\usepackage[latin1]{inputenc}
\markboth{\centerline{\textsl{\bfseries Draft of QI open problem -- content test only!}}}{\centerline{\textsl{\bfseries Draft of QI open problem -- content test only!}}}
\pagestyle{myheadings}
\title{Draft of QI open problem -- content test only!}
\author{No page formatting}
\fi

%
% ***************************************************************************
%
% Praeambel, redigiert durch IMaPh
%

% Keine AMS-Erweiterungen!
% Standardbefehle (\text{}, \R, \C) sind bereits in qiheader.tex
% definiert.

% TTH fuegt Label "BEGINDOCUMENT" ein und ignoriert den Block
%%tth: \iftth BEGINDOCUMENT \begin{document} \else
\ifx\quantph\undefined
	\begin{document}
\fi
\maketitle
% \makeinfo
%%tth: \fi

%
% ***************************************************************************
%
% Text-Teil, redigiert durch IMaPh
%

%
% NICHT VERGESSEN:
%	- ggf. $...$ durch \(...\) ersetzen
%	- Eintraege ins Literaturverzeichnis vollstaendig klammern: \bibitem[ABC]{25:ABC}{A.A.,B.B., ...}
%

\section{Problem}

Are two-way distillible entanglement and secret key rate lockable?

\section{Background}

An entanglement measure is lockable, if it is
extremely sensitive to the loss of a single qubit by one of the
partners \cite{25:HHHO04}. In this paper it is shown that entanglement of
formation, entanglement cost, logarithmic negativity, and all
convex, asymptotically discontinuous entanglement measures are
lockable. More recently \cite{25:CW05}, also squashed entanglement has been
shown to be lockable.

\ifx\quantph\undefined
	\end{document}
\fi

%\newpage\centerline{}

%
% Schalter für End-Version: \let\final\relax
% Schalter für Vorschau-Version: \final nicht definiert
%
\ifx\final\relax

%
% Header-Datei laden (voller Pfad, kein Endung ".tex")
%

\ifx\quantph\undefined
\else
% Automatische Definition der Variablen
\qpvar{\qpno}
\qpvar{\qptitle}
\qpvar{\qpdate}
\qpvar{\qpcontribute}
\qpvar{\qpfirstdate}
\qpvar{\qplastdate}
\qpvar{\qpsolve}
\fi

%
% ***************************************************************************
%
% Variablenbelegungen in der LaTeX-Datei,
% einzufuegen durch IMaPh
%
% Diese Variablen sollten NICHT im TEXT-Teil
% definiert oder verwendet werden!
%

% Der folgende Block wird von TTH "verbatim" ausgegeben,
% alle anderen verarbeiten ihn
%%tth: \begin{html}
\qpno{26}
\qptitle{Bell inequalities holding for all quantum states}
\qpdate{11 Apr 2005}
\qpcontribute{\href{http://www.imaph.tu-bs.de/qi/links.html}{R. Gill}}
\qpfirstdate{11 Apr 2005}
\qplastdate{\nothing}
\qpsolve{\nothing}
%%tth: \end{html}

%
% Für Vorschau-Version nur das nötigste einstellen
%
% \nonstopmode kann ggf. auskommentiert werden!
%
\else
\nonstopmode
\documentclass[12pt,a4paper]{article}
\usepackage[latin1]{inputenc}
\markboth{\centerline{\textsl{\bfseries Draft of QI open problem -- content test only!}}}{\centerline{\textsl{\bfseries Draft of QI open problem -- content test only!}}}
\pagestyle{myheadings}
\title{Draft of QI open problem -- content test only!}
\author{No page formatting}
\fi

%
% ***************************************************************************
%
% Praeambel, redigiert durch IMaPh
%

% Keine AMS-Erweiterungen!
% Standardbefehle (\text{}, \R, \C) sind bereits in qiheader.tex
% definiert.
\providecommand{\GL}{\mathrm{GL}}

%-ltoh-	:{}		:\subsection[*]?:<br><p><font size=+1><b>:</b></font>:

% TTH fuegt Label "BEGINDOCUMENT" ein und ignoriert den Block
%%tth: \iftth BEGINDOCUMENT \begin{document} \else
\ifx\quantph\undefined
	\begin{document}
\fi
\maketitle
% \makeinfo
%%tth: \fi

%
% ***************************************************************************
%
% Text-Teil, redigiert durch IMaPh
%

%
% NICHT VERGESSEN:
%	- ggf. $...$ durch \(...\) ersetzen
%	- Eintraege ins Literaturverzeichnis vollstaendig klammern: \bibitem[ABC]{26:ABC}{A.A.,B.B., ...}
%

\section{Problem}

The setting for this problem is the same as for \href{http://www.imaph.tu-bs.de/qi/problems/1.html}{Problem 1}: We
consider correlations between $N$ parties, each of which can
perform $M$ different measurements yielding one of $K$ possible
outcomes each. We can reduce the number of dimensions by
considering only those correlation data satisfying the
no-signalling constraint, i.\,e., the choice of a measuring device
by one party $A$ never changes the (joint) probabilities seen by
all the other parties, unless results are selected with respect to
the outcomes of $A$. Only obeying no-signalling and positivity
constraints, we get the \emph{no-signalling polytope} $P$.
Contained in it is the convex body $Q$ of correlations obtainable
from a multipartite quantum state with \emph{quantum mechanical POVM
measurements}, and inside $Q$ the polytope $C$ of correlations
realizable by a \emph{classical realistic theory} (see Figure).

%%tth:  \iftth
%%tth:  \special{html:<img src="fig26-1.png">}
%%tth:  \else
%-ltoh- +{2}+\begin#1#2+<img src="fig26-1.png"><p>++
%-ltoh- :comm:\end{figure}:::

\begin{figure}[h]{
	\begin{center}
	\includegraphics[width=0.5\textwidth]{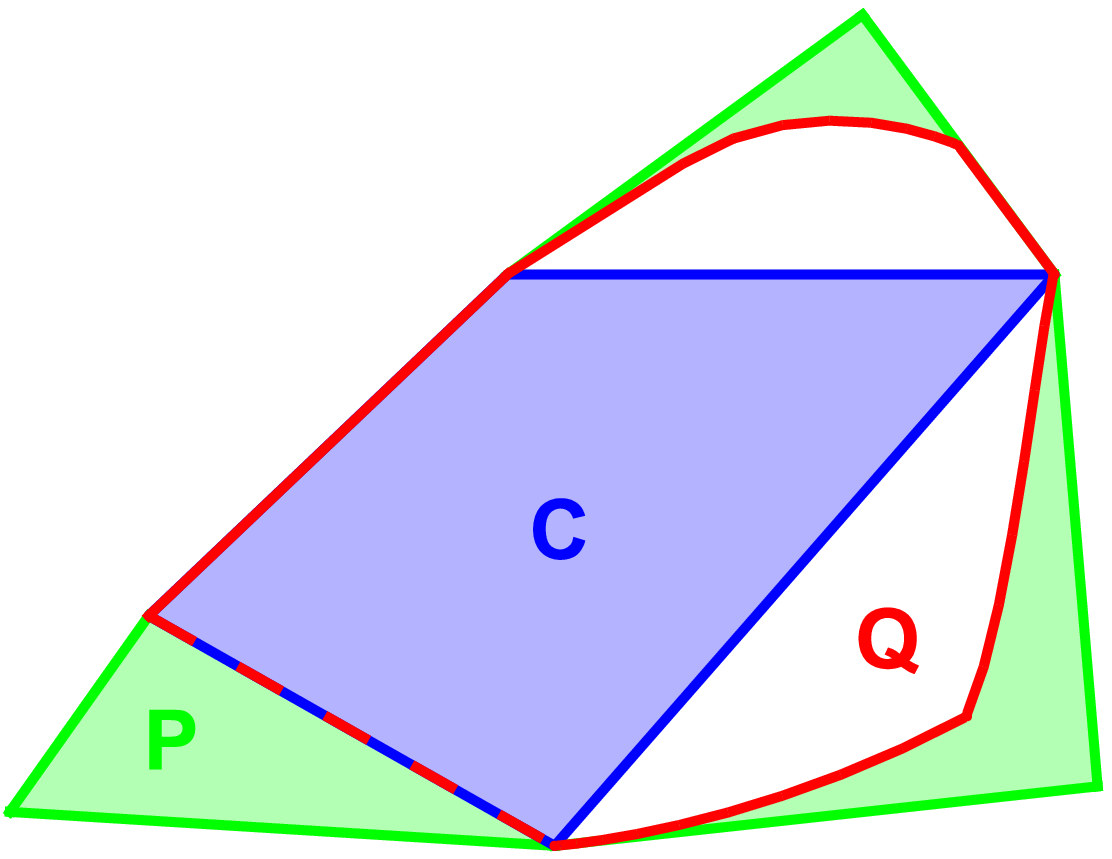}
	\end{center}
}\end{figure}

%%tth:  \fi

\newpage

Here are some questions about the way $Q$ fits in between the
polytope $P$ and $C$:

\subsection*{Problem 26.A:}

Consider the part of the boundary of $Q$, which is not already contained in the
boundary of $P$. Can one reach all these points by choosing each one of the local
Hilbert spaces to be $K$-dimensional, and each measurement as a complete von
Neumann measurement (with $K$ orthogonal projectors) on pure states with minimal
dimension?

\subsection*{Problem 26.B:}

Consider a maximal face of the polytope $C$, which is not also a face of $P$ (a
blue line in the above figure). In other words, consider a "proper Bell
inequality", i.\,e., a tight linear inequality for local classical correlations,
which does not follow from positivity and no-signalling. Then can we find points
of $Q$ outside the face? Or, phrased in terms of Bell inequalites, can every
proper Bell inequality be violated by quantum correlation data? In the above
figure, this asks whether or not a face like the dashed red/blue line can occur.

\ifx\quantph\undefined
	\end{document}
\fi

\addtocontents{toc}{\protect\end{tabular}}
\addtocontents{toc}{\protect\clearpage\protect\noindent}
\addtocontents{toc}{\protect\begin{tabular}{@{}c@{}c@{}c@{}c@{}>{\protect\centering}p{25\cu}@{}c@{}}}
\addtocontents{toc}{\protect\tocback{No}&\protect\multicolumn{4}{l}{\tocback{Title}}&\protect\tocback{Page}\\%
	&\protect\tocback{Contact}&\protect\tocback{Date}&\protect\tocback{Last progress}&\protect\tocback{Solved by}%
	\protect\tabularnewline[2ex]}
\addtocontents{toc}{\hspace*{5\cu}&\hspace*{27\cu}&\hspace*{18\cu}&\hspace*{18\cu}&\hspace*{27\cu}&\hspace*{5\cu}\\}

%
% Schalter für End-Version: \let\final\relax
% Schalter für Vorschau-Version: \final nicht definiert
%
\ifx\final\relax

%
% Header-Datei laden (voller Pfad, kein Endung ".tex")
%

\ifx\quantph\undefined
\else
% Automatische Definition der Variablen
\qpvar{\qpno}
\qpvar{\qptitle}
\qpvar{\qpdate}
\qpvar{\qpcontribute}
\qpvar{\qpfirstdate}
\qpvar{\qplastdate}
\qpvar{\qpsolve}
\fi

%
% ***************************************************************************
%
% Variablenbelegungen in der LaTeX-Datei,
% einzufuegen durch IMaPh
%
% Diese Variablen sollten NICHT im TEXT-Teil
% definiert oder verwendet werden!
%

% Der folgende Block wird von TTH "verbatim" ausgegeben,
% alle anderen verarbeiten ihn
%%tth: \begin{html}
\qpno{27}
\qptitle{The power of CGLMP inequalities}
\qpdate{15 Apr 2005}
\qpcontribute{\href{http://www.imaph.tu-bs.de/qi/links.html}{R. Gill}}
\qpfirstdate{15 Apr 2005}
\qplastdate{\nothing}
\qpsolve{\nothing}
%%tth: \end{html}

%
% Für Vorschau-Version nur das nötigste einstellen
%
% \nonstopmode kann ggf. auskommentiert werden!
%
\else
\nonstopmode
\documentclass[12pt,a4paper]{article}
\usepackage[latin1]{inputenc}
\markboth{\centerline{\textsl{\bfseries Draft of QI open problem -- content test only!}}}{\centerline{\textsl{\bfseries Draft of QI open problem -- content test only!}}}
\pagestyle{myheadings}
\title{Draft of QI open problem -- content test only!}
\author{No page formatting}
\fi

%
% ***************************************************************************
%
% Praeambel, redigiert durch IMaPh
%

% Keine AMS-Erweiterungen!
% Standardbefehle (\text{}, \R, \C) sind bereits in qiheader.tex
% definiert.
\providecommand{\GL}{\mathrm{GL}}

%-ltoh-	:{}		:\subsection[*]?:<br><p><font size=+1><b>:</b></font>:

% TTH fuegt Label "BEGINDOCUMENT" ein und ignoriert den Block
%%tth: \iftth BEGINDOCUMENT \begin{document} \else
\ifx\quantph\undefined
	\begin{document}
\fi
\maketitle
% \makeinfo
%%tth: \fi

%
% ***************************************************************************
%
% Text-Teil, redigiert durch IMaPh
%

%
% NICHT VERGESSEN:
%	- ggf. $...$ durch \(...\) ersetzen
%	- Eintraege ins Literaturverzeichnis vollstaendig klammern: \bibitem[ABC]{27:ABC}{A.A.,B.B., ...}
%

\section{Problem}

In the setting of \href{http://www.imaph.tu-bs.de/qi/problems/26.html}{problem 26},
consider especially the case $(N,M,K)=(2,2,d)$.

\subsection*{Problem 27.A:}

   Show that every face of the local polytope $C$, which is not
   already contained in a face of the no-signalling polytope $P$ is
   of CGLMP type, i.\,e., an inequality of the form first written out
   in \cite{27:CGLMP}, but possibly lifted from lower dimensions by fusing
   together some outcomes.

\subsection*{Problem 27.B:}

   Numerically, the observables maximally violating the CGLMP inequality on a
   maximally entangled state are of a very specific form \cite{27:DKZ}, involving
   measurements in computational basis, transformed by only discrete Fourier
   transformation and diagonal unitaries \cite{27:CGLMP}. Show that this is
   necessarily the case. Show also that these measurements realize the highest
   resistance of violation to noise, and the best discrimination against classical
   realism in the sense of Kullback-Leibler divergence \cite{27:Gill1}.

\section{Background}

%-ltoh-	:{}		:\mathsf:<b>:</b>:
%-ltoh-	:comm	:\mod:mod::
%%tth:	\renewcommand{\mathsf}{\mathbf}
%%tth:	\newcommand{\mod}{mod }

According to the setting $(N,M,K)=(2,2,d)$, the CGLMP inequality features two
parties, $X$ and $Y$, with two observables each: $X_1, X_2$ and $Y_1, Y_2$,
respectively. Each observable has $d$ possible outcomes. In order to simplify
notation, we use the function $m(x) = x \mod d$ where $m(x)\in \{0,1,...,d-1\}$
for integer $x$ and we denote expectation values by $\mathsf{E}$. The inequality
can then be written \cite{27:Gill2}:
\[
	\mathsf{E}(m(X_1-Y_1)) + \mathsf{E}(m(Y_1-X_2)) + \mathsf{E}(m(X_2-Y_2)) + 
		\mathsf{E}(m(Y_2-X_1-1)) \ge d-1.
\]
This statement also suggests a very elegant proof of the inequality \cite{27:Gill2}:
Note that $(X_1-Y_1) + (Y_1-X_2) + (X_2-Y_2) + (Y_2-X_1-1) = -1$. Apply the
function $m$ to both sides, and use $m(a)+m(b)+m(c)+m(d) \ge m(a+b+c+d)$.

\ifx\quantph\undefined
	\end{document}
\fi

%
% Schalter für End-Version: \let\final\relax
% Schalter für Vorschau-Version: \final nicht definiert
%
\ifx\final\relax

%
% Header-Datei laden (voller Pfad, kein Endung ".tex")
%

\ifx\quantph\undefined
\else
% Automatische Definition der Variablen
\qpvar{\qpno}
\qpvar{\qptitle}
\qpvar{\qpdate}
\qpvar{\qpcontribute}
\qpvar{\qpfirstdate}
\qpvar{\qplastdate}
\qpvar{\qpsolve}
\fi

%
% ***************************************************************************
%
% Variablenbelegungen in der LaTeX-Datei,
% einzufuegen durch IMaPh
%
% Diese Variablen sollten NICHT im TEXT-Teil
% definiert oder verwendet werden!
%

% Der folgende Block wird von TTH "verbatim" ausgegeben,
% alle anderen verarbeiten ihn
%%tth: \begin{html}
\qpno{28}
\qptitle{Local equivalence of graph states}
\qpdate{21 Apr 2005}
\qpcontribute{\href{http://www.imaph.tu-bs.de/staff/dsl.html}{D. Schlingemann}}
\qpfirstdate{21 Apr 2005}
\qplastdate{\nothing}
\qpsolve{\nothing}
%%tth: \end{html}

%
% Für Vorschau-Version nur das nötigste einstellen
%
% \nonstopmode kann ggf. auskommentiert werden!
%
\else
\nonstopmode
\documentclass[12pt,a4paper]{article}
\usepackage[latin1]{inputenc}
\markboth{\centerline{\textsl{\bfseries Draft of QI open problem -- content test only!}}}{\centerline{\textsl{\bfseries Draft of QI open problem -- content test only!}}}
\pagestyle{myheadings}
\title{Draft of QI open problem -- content test only!}
\author{No page formatting}
\fi

%
% ***************************************************************************
%
% Praeambel, redigiert durch IMaPh
%

% Keine AMS-Erweiterungen!
% Standardbefehle (\text{}, \R, \C) sind bereits in qiheader.tex
% definiert.

% TTH fuegt Label "BEGINDOCUMENT" ein und ignoriert den Block
%%tth: \iftth BEGINDOCUMENT \begin{document} \else
\ifx\quantph\undefined
	\begin{document}
\fi
\maketitle
% \makeinfo
%%tth: \fi

%
% ***************************************************************************
%
% Text-Teil, redigiert durch IMaPh
%

%
% NICHT VERGESSEN:
%	- ggf. $...$ durch \(...\) ersetzen
%	- Eintraege ins Literaturverzeichnis vollstaendig klammern: \bibitem[ABC]{28:ABC}{A.A.,B.B., ...}
%

\section{Problem}

Decide whether two graph states, which can be mapped into each
other by a local unitary, can also be mapped into each other by a
local unitary from the Clifford group.

\section{Background}

\emph{Graph states} \cite{28:Schl} are multiparticle states which are
associated with graphs. Each vertex of the graph corresponds to a
qubit. The links describe contributions to the phase of the vector
components in computational basis:
\[
	\langle q_1,q_2,\cdots,q_n|\psi\rangle
		=2^{-n/2} \prod_{\text{edges $i,j$}} (-1)^{q_i q_j},
\]
where each $q_i=0,1$. They can also be characterized
\cite{28:HEB04} by eigenvalue equations of stabilizer form
\[
	X^{(i)} \prod_{j:\,\text{edge $i,j$}} Z^{(j)}\quad\psi=\psi,
\]
where $X^{(i)}$ and $Z^{(i)}$ stand for the x- and z-Pauli matrices at vertex $i$.
The Pauli operators which leave the vector $\psi$ invariant generate the
\emph{stabilizer group} of the graph state.  

The \emph{Clifford group} consists of those unitary operators $U$,
such that $UPU^*$ is a multiple of a Pauli matrix, whenever $P$ is
a Pauli matrix. This group is generated by the Hadamard matrix 
\[
	1/\sqrt{2}\left(\begin{array}{cc}
	1 & 1 \\
	1 & -1
	\end{array}\right)
	\text{, the diagonal matrix }
	\left(\begin{array}{cc}
	1 & 0 \\
	0 & i
	\end{array}\right)
\]
and the Pauli matrices themselves.

The notion of graph states and the Clifford group can be
generalized in a natural way to non-binary systems \cite{28:Schl}.

\section{Partial results}

A complete set of invariants for locally unitary equivalent qubit graph states is
given by Van~den~Nest et~al. \cite{28:NDM04b}. A complete set of invariants
of the local Clifford group is also known \cite{28:NDM04d,28:NDM04e}.
However, the relation between these two invariants is still unclear. At least for
the qubit case it is conjectured that a complete family of local Clifford
invariants exists which is contained in the class of local unitary invariants.    

It has been shown by Van~den~Nest et~al. \cite{28:NDM04c} that for a
particular class of qubit graph states local unitary equivalence implies local
Clifford equivalence. This class consists of graph states for which stabilizer
group $S$ has a particular structure. For instance, all stabilizer states that can
be derived from a $GL(4)$-linear code belong to this class.  

Examples which do not belong to this class are generalized GHZ states that
correspond to star shaped graphs, i.\,e. one of the qubits is connected with each
of the remaining qubits. Nevertheless it has been shown in \cite{28:NDM04c}
that local unitary equivalence implies local Clifford equivalence also in this
case. 

As outlined by Hein et~al. \cite{28:HEB04}, numerical results show that local
Clifford equivalence coincides with local unitary equivalence for qubit graph
states associated with connected graphs up to 7 vertices.

\ifx\quantph\undefined
	\end{document}
\fi

%\newpage\centerline{}

%
% Schalter für End-Version: \let\final\relax
% Schalter für Vorschau-Version: \final nicht definiert
%
\ifx\final\relax

%
% Header-Datei laden (voller Pfad, kein Endung ".tex")
%

\ifx\quantph\undefined
\else
% Automatische Definition der Variablen
\qpvar{\qpno}
\qpvar{\qptitle}
\qpvar{\qpdate}
\qpvar{\qpcontribute}
\qpvar{\qpfirstdate}
\qpvar{\qplastdate}
\qpvar{\qpsolve}
\fi

%
% ***************************************************************************
%
% Variablenbelegungen in der LaTeX-Datei,
% einzufuegen durch IMaPh
%
% Diese Variablen sollten NICHT im TEXT-Teil
% definiert oder verwendet werden!
%

% Der folgende Block wird von TTH "verbatim" ausgegeben,
% alle anderen verarbeiten ihn
%%tth: \begin{html}
\qpno{29}
\qptitle{Entanglement of formation for Gaussian states}
\qpdate{21 Apr 2005}
\qpcontribute{\href{http://www.imaph.tu-bs.de/staff/ok.html}{O. Krüger}}
\qpfirstdate{21 Apr 2005}
\qplastdate{\nothing}
\qpsolve{\nothing}
%%tth: \end{html}

%
% Für Vorschau-Version nur das nötigste einstellen
%
% \nonstopmode kann ggf. auskommentiert werden!
%
\else
\nonstopmode
\documentclass[12pt,a4paper]{article}
\usepackage[latin1]{inputenc}
\markboth{\centerline{\textsl{\bfseries Draft of QI open problem -- content test only!}}}{\centerline{\textsl{\bfseries Draft of QI open problem -- content test only!}}}
\pagestyle{myheadings}
\title{Draft of QI open problem -- content test only!}
\author{No page formatting}
\fi

%
% ***************************************************************************
%
% Praeambel, redigiert durch IMaPh
%

% Keine AMS-Erweiterungen!
% Standardbefehle (\text{}, \R, \C) sind bereits in qiheader.tex
% definiert.
\providecommand{\GL}{\mathrm{GL}}

%-ltoh-	:{}		:\subsection[*]?:<br><p><font size=+1><b>:</b></font>:

% TTH fuegt Label "BEGINDOCUMENT" ein und ignoriert den Block
%%tth: \iftth BEGINDOCUMENT \begin{document} \else
\ifx\quantph\undefined
	\begin{document}
\fi
\maketitle
% \makeinfo
%%tth: \fi

%
% ***************************************************************************
%
% Text-Teil, redigiert durch IMaPh
%

%
% NICHT VERGESSEN:
%	- ggf. $...$ durch \(...\) ersetzen
%	- Eintraege ins Literaturverzeichnis vollstaendig klammern: \bibitem[ABC]{29:ABC}{A.A.,B.B., ...}
%

\section{Problem}

Entanglement of formation is defined as a minimimum over all convex decompositions
of a bipartite state into pure states (see
\hyperlink{Outline7.0}{problem 7}). It has been
shown  for certain two-mode Gaussian states this minimum can be taken over
decompositions of the given state into pure states, all of which are translates of
the same squeezed Gaussian state, with Gaussian weights.

Show (or disprove) that this is true for all Gaussian states.

\section{Background}

If the optimization over convex decompositions of a bipartite state is restricted
to decompositions into Gaussian states, entanglement of formation becomes a new
entanglement measure, the \emph{Gaussian entanglement of formation} introduced in
\cite{29:1}. With this, the above question reads: Does Gaussian entanglement of 
formation equal entanglement of formation for all Gaussian states?

\section{Partial Results}

It has been shown in \cite{29:2} that Gaussian entanglement of formation equals
entanglement of formation for two-mode Gaussian states which are symmetric with
respect to interchange of the modes.

\ifx\quantph\undefined
	\end{document}
\fi

\addtocontents{toc}{\protect\end{tabular}}
\addtocontents{toc}{\protect\enlargethispage{1ex}}


\begin{thebibliography}{BBPS96}
\bibitem[BFC]{20:BFC}{This was boldly conjectured by the author and is in certain
	circles known as the Big-Fat-Conjecture.}
\bibitem[Bet]{20:Bet}{The existence of such a class is the subject of a
	bet between Michal Horodecki and Reinhard Werner.}
\end{thebibliography}
\end{document}